\documentclass[12pt,aps,tightenlines,notitlepage]{revtex4-2}
\usepackage{amsmath}
\usepackage{amssymb}
\usepackage{bm}
\usepackage[mathscr]{eucal}
\usepackage{graphicx}
\usepackage{psfrag}
\usepackage{subfigure}
\usepackage{color}
\usepackage{wasysym}
\include{graphix}
\usepackage{framed}
\usepackage{enumitem}%
\usepackage{lipsum}
\usepackage{float}
\usepackage{stmaryrd}
\usepackage{MnSymbol}
\definecolor{orange}{rgb}{1,0.5,0}
\usepackage[normalem]{ulem}
\usepackage{cancel}
\usepackage{hyperref}

\usepackage{xcolor}
\usepackage{listings}

\definecolor{bg}{RGB}{255,255,226}

\usepackage{tikz}
\usepackage{pgfplots}

\usetikzlibrary{patterns}
\usetikzlibrary{shapes}

\graphicspath{{figs/}}

\newenvironment{remark}[1][Remark]{\begin{trivlist}
\item[\hskip \labelsep {\bfseries #1}]}{\end{trivlist}}

\newcommand{\myqed}{\nobreak \ifvmode \relax \else
      \ifdim\lastskip<1.5em \hskip-\lastskip
      \hskip1.5em plus0em minus0.5em \fi \nobreak
      \vrule height0.75em width0.5em depth0.25em\fi}

\newcommand{\mathd}{\mathrm{d}}
\newcommand{\imag}{\mathrm{i}}
\newcommand{\mathe}{\mathrm{e}}

\newcommand{\vecu}{\bm{u}}
\newcommand{\vecx}{\bm{x}}
\newcommand{\vecF}{\bm{F}}

\newcommand{\hatn}{\widehat{\bm{n}}}
\newcommand{\surften}{\gamma}
\newcommand{\hatz}{\widehat{\mathbf{z}}}

\newcommand{\phihatn}{\widehat{\phi}_n}
\newcommand{\phihatzero}{\widehat{\phi}_0}
\newcommand{\xiL}{f_0}
\newcommand{\omegalin}{\Omega_{\mathcal{L}}}

\begin{document}

\title{An integrated theoretical, experimental, and numerical study of small-amplitude water waves}
\author{Lennon \'O N\'araigh$^1$, Nicolas Farault$^{1,2}$, Nicola Young$^1$ }
\email{onaraigh@maths.ucd.ie}
\address{$^1\,$School of Mathematics and Statistics, University College Dublin, Belfield, Dublin 4, Ireland}
\address{$^2\,$Polytech Lyon, 15, Boulevard Latarjet 69622 VILLEURBANNE CEDEX}
\date{\today}

\begin{abstract}
We introduce an inexpensive experimental setup for analyzing free-surface water waves: a $1\,\mathrm{m}$-long tabletop flume made from perspex, driven by a variable-frequency piston wavemaker built from Lego.   Using mobile-phone video capture, we collect experimental data and compare it with predictions from linear gravity-capillary wave theory and with multiphase simulations performed in OpenFOAM. We find excellent quantitative agreement across all three approaches.  Our setup
may be valuable for students with  a background in Mathematical Modelling who lack hands-on laboratory experience.  To explore this, we report on a survey of students who completed an integrated theoretical, experimental, and computational project. While students found the experience enhanced their learning of Fluid Mechanics, they also noted the need for better support in setting up and running CFD simulations.
\end{abstract}

\maketitle

\section{Introduction}
\label{sec:intro}

The linear theory of water waves is typically introduced to Physics, Engineering, and Applied Mathematics students after more introductory modules have been completed in Calculus of Several Variables, Vector Calculus, and Mathematical Methods.  For many students taking a more theoretically-oriented programme of study, the theory remains just that, and students' theoretical understanding is often not matched by a practical lab-based experience.  Equally, the simulation of water waves requires the simulation of a multiphase flow problem, which requires sophisticated interface-capturing algorithms often beyond the scope of a typical undergraduate programme.  In this paper, we report on an instructional framework which combines theoretical modelling, computational modelling using ready-made OpenFOAM case studies, and experiments using a  practical, inexpensive flume with a custom-made wavemaker, made from Lego, thereby filling in gaps in students' learning.  

\subsection{Literature Review}

The study of water waves is important for its practical applications in wave forecasting, coastal engineering~\cite{madsen1970waves}, and for tsunami warnings.  In a first approximation, linear theory is applied, wherein the wave amplitude is small compared to the water depth.  Then, the relevant equations of motion (whether the Euler equations for an inviscid fluid, or the Navier--Stokes equations for a viscous one) simplify into linear partial differential equations.  Conditions at the interface typically involving the kinematic condition (namely that the interface moves with the fluid) are then used to close the equations.  In many textbooks~\cite{feynman1965feynman,acheson1990elementary,jog2015fluid}, a temporal analysis is performed, wherein an initial sinusoidal disturbance on the interface is introduced with wavenumber $k$.  This is done via a prescribed initial condition.  Then, the linear equations of motion reduce to an eigenvalue problem, in which the wave frequency is obtained as a function of $k$, which gives rise to the dispersion relation, $\omega=\omega(k)$.  The dispersion relation may depend on other fluid parameters as well.  For instance, for water waves in a fluid of finite depth $h$, the standard dispersion relation is~\cite{acheson1990elementary}:
\begin{equation}
\omega(k)=\left[k\left(g+\frac{\surften}{\rho}k^2\right)\tanh(kh)\right]^{1/2},
\label{eq:disp1}
\end{equation}
where $g$ is the acceleration due to gravity, $\rho$ is the liquid density, and $\surften$ is the surface tension.  

A more advanced analysis is concerned with a time-dependent localized disturbance is introduced to the system, and the downstream propagation of the disturbance is considered.  Such spatio-temporal analysis forms the basis of wavemaker theory~\cite{dean1991water}, and is also important in understanding absolute and convective instability in more advanced problems~\cite{huerre1990local,naraigh2013absolute}.  A spatio-temporal formulation of the water-wave problem can help to bridge the gap to these research-level problems, and is therefore introduced in this paper.

In order to validate the dispersion relation~\eqref{eq:disp1} (or its spatio-temporal analogue), researchers  carry out experiments in a wave tank.  Referring back to Equation~\eqref{eq:disp1}, gravity waves refer to waves for which $g\gg (\surften/\rho)k^2$ (hence, long waves), and these are often of most interest to researchers due to the aforementioned applications.  To observe such waves in an experiment, long channels are required.  To ensure unidirectional wave propagation, a long narrow channel is used, such that the waves are two-dimensional.  Such a setup is referred to as a flume.  One such example is the laboratory-scale flume at the University of Warwick~\cite{dong2024improved} with dimensions $22\,\mathrm{m}$ long,  $0.6\,\mathrm{m}$ wide, and $1.0\,\mathrm{m}$ deep.   In contrast, capillary waves are short and occur when $g\ll (\surften/\rho)k^2$.  Such capillary waves appear as nonlinear, secondary disturbances which occur as a perturbation with respect to a primary gravity wave.  They are also of interest in the classroom setting in so-called ripple tanks, which are small (typically, less than $0.5\,\mathrm{m}\times 0.5\,\mathrm{m}$ and as such, support only capillary waves~\cite{ripple}.  In this work, we document the construction of a tabletop flume, of length $1\,\mathrm{m}$, which supports waves for which both gravity and surface tension play an important role in the dispersion relation.

%

Numerical simulations also provide important insights into water waves.  In the first instance, they can be used to validate the linear theory.  They provide a surrogate for experiments, such that the flow generated by the interfacial waves can be inferred, without having to resort to experimental techniques such as PIV.  Finally, simulations can be used to provide insights into nonlinear waves, including wave breaking and wave overturning, which cannot be described by linear theory.  While classical water-wave theory is inviscid and relies on the Euler equations, many computational fluid dynamics (CFD) software frameworks rely on the Navier--Stokes equations, which include the effect of viscosity.  However, provided the viscosity is small (in a sense to be determined), water-wave modelling with the Navier--Stokes equations will provide the same answer was water-wave modelling with the Euler equations.


\subsection{Aim of the paper}

The aim of the paper is to introduce students and instructors to an inexpensive piece of apparatus and complementary open-source fluid simulation software, which 
together provide a hands-on way to investigate  the linear theory of free-surface waves. Crucially, with high-quality video recording now available on most mobile phones, students already carry a powerful data acquisition tool in their pockets. We show how this  technology can be used to capture wave motion in the flume, enabling direct comparison between experimental observations, theoretical predictions, and numerical simulations, thereby making a complex topic more tangible for students.

\subsection{Plan of the paper}

The theoretical spatio-temporal analysis is set out in Section~\ref{sec:theory} for open wave tanks and in Section~\ref{sec:theory_closed} for closed wave tanks.  The open wave tank admits travelling waves whereas the closed wave tank admits a standing wave.  As a by-product, this analysis provides a derivation of Equation~\eqref{eq:disp1}.  To illustrate the theory, experiments using a tabletop flume are carried out in Section~\ref{sec:flume}. 
We carry out statistical analysis on the data emanating from the experiments to show that the observed wave pattern is a linear combination of travelling waves and standing waves.
 To illustrate the wave phenomena in more detail, and to provide a rigorous test of the dispersion relation~\eqref{eq:disp1}, computational fluid dynamics simulations are presented in Section~\ref{sec:cfd}.  The experience of students using this three-fold approach to learning about water waves is described in Section~\ref{sec:feedback}.  Concluding remarks are presented in Section~\ref{sec:conc}.

\section{Spatio-temporal analysis of small-amplitude water waves: the open tank}
\label{sec:theory}

In this section we develop the spatio-temporal theory of small-amplitude water waves, for an open tank.     This theory describes the linear response of the free surface to a localized forcing corresponding to a wavemaker and as such, forms the basis of wavemaker theory.  The theory has already been presented in the standard reference~\cite{dean1991water} and is included here for completeness, and to provide the proper context for the subsequent experimental and computational investigations.

\begin{figure}
\centering
\begin{tikzpicture}[scale=1.2,transform shape]
    \draw[->] (0,0) -- (0,2) node[above] {$z$};

    \draw[thick] (0,0) -- (8,0); 
    
	\draw[-] (-0.5,1) -- (0,1);
    \draw[-,dashed] (0,1) -- (8,1); 
	\draw[->] (8,1)--(8.5,1);
	\node[right] at (8.5,1) {$x$};

    \draw[thick,blue,smooth,samples=100,domain=0.2:8] 
        plot(\x,{1+0.1*sin(2*pi*\x r)});
    
    \fill[gray!50] (-0.2,0) rectangle (0.2,1.5);
    \draw[thick] (-0.2,0) -- (-0.2,1.5);
    \draw[thick] (0.2,0) -- (0.2,1.5);
	\draw[thick] (-0.2,0) -- (0.2,0);
    
    \draw[->,thick] (0.2,0.5) -- (0.4,0.5) node[right] {$\xi(z,t)$};
    \draw[->,thick] (-0.2,0.5) -- (-0.4,0.5);

    \draw[->,thick] (7,2.5) -- (7,1.8) node[midway,right] {$g$};

    \node at (3.5,1.5) {Free Surface: $z=\eta(x,t)$};
    \node at (0,-0.5) {$x=0$};
		
	\node at (-1,1) {$z=0$};
	\node at (-1,0) {$z=-h$};
	
	\node at (6,0.5) [rectangle,draw] {$\Omega$};
\end{tikzpicture}
\caption{Schematic diagram showing the generation of small-amplitude water waves by a piston wavemaker located at $x=0$ (open tank)}
\label{fig:schematic1}
\end{figure}
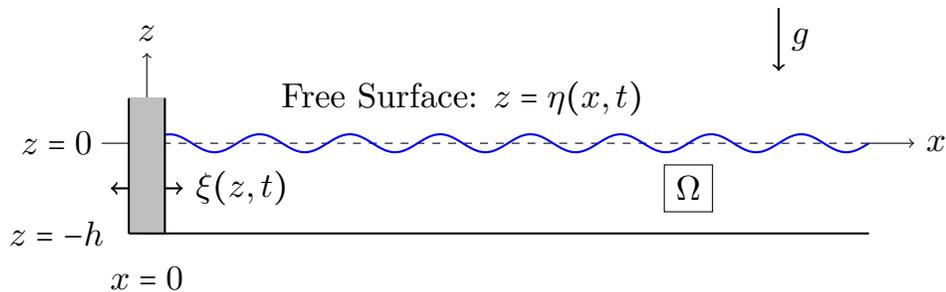

For this purpose, we refer to the set-up in Figure~\ref{fig:schematic1}, and take the direction of propagation along the $x$-axis, and the direction of oscillation along the $z$-axis.  The figure describes an open tank, in which $x\in [0,\infty)$.
  The free surface is therefore denoted by $z=\eta(x,t)$, where $z=0$ represents the undisturbed free-surface height.  Standard undergraduate texts describe a temporal theory~\cite{acheson1990elementary}, where the free surface is initialized to have a monochromatic sinusoidal profile $\eta(x,t=0)\propto \sin(kx+\varphi)$ everywhere (here, $\varphi$ is a constant phase term).  Here, we describe in detail the spatio-temporal theory, wherein the free surface is assumed to be undisturbed initially, but to undergo a localized forcing at $x=0$ corresponding to the impact of a piston wavemaker.

%
%

To understand the setup of the spatio-temporal wave propagation, we refer to Figure~\ref{fig:schematic1}.  A piston located at $x=0$ generates localized, impulsive forcing.  The piston oscillates  according to:
\begin{equation}
\xi(z,t)=\Re\left[-\frac{1}{\imag\omega}f(z)\mathe^{-\imag \omega t}\right].
\label{eq:piston}
\end{equation}
where $f(z)$ is a shape function describing the details of the back-and-forth motion of the piston.  This can be left unspecified for now.
Inside the domain $\Omega$, the flow is inviscid and irrotational, so potential theory applies:
\begin{equation}
\nabla^2\Phi=0,\qquad \vecx\in \Omega.
\end{equation}
Here, $\Phi$ is the velocity potential, such that $\vecu=\nabla\Phi$.  Also, the vector $\vecx=(x,z)$ is a two-dimensional vector.
The boundary condition at $z=-h$ is the no-penetration condition, $w=0$, hence:
\begin{equation}
\frac{\partial\Phi}{\partial z}=0,\qquad z=-h.
\end{equation}

\subsection{Conditions at the free surface}

We next look at the boundary condition at the free surface $z=\eta$.  Bernoulli's equation gives the pressure on the free surface as:
\begin{equation}
p=-\rho \frac{\partial\phi}{\partial t}-\tfrac{1}{2}\rho \vecu^2-\rho g \eta+f(t),
\end{equation}
where $f(t)$ is a parameter associated with Bernoulli's principle. 
We assume that the wave amplitude is small in comparison to the water depth $h$.  This introduces a small parameter $\epsilon=\max(\eta)/h$ into the problem.  Thus, disturbances, whether of amplitude, pressure, velocity or streamfunction are proportional to $\epsilon$, whereas products of disturbances (such as $\vecu^2$) are proportional to $\epsilon^2$ and can be neglected in a  small-amplitude approximation.  Thus, the pressure on the free surface can be approximated as:
\begin{equation}
p=-\rho \frac{\partial\Phi}{\partial t}-\rho g \eta+f(t).
\label{eq:bernoulli}
\end{equation}
From Reference~\cite{acheson1990elementary}, the pressure condition at the interface for an inviscid flow is:
\begin{equation}
p_{atm}-p=\gamma \kappa,
\end{equation}
%
%
where $\kappa=\eta_{xx}/(1+\eta_x^2)^{3/2}$ is the mean curvature and $p_{atm}$ is the atmospheric pressure.  In the small-amplitude approximation, we have:
\begin{equation}
p_{atm}-p=\gamma \eta_{xx}.
\end{equation}
%
%
Using Equation~\eqref{eq:bernoulli}, this becomes:
\begin{equation}
\rho \frac{\partial\Phi}{\partial t}+\rho g \eta+\left[p_{atm}-f(t)\right]=\surften\eta_{xx},\qquad z=\eta.
\end{equation}
Since $f(t)$ is arbitrary, we set $f(t)=p_{atm}$, leaving:
\begin{equation}
\rho\frac{\partial\Phi}{\partial t}+\rho g\eta=\surften\eta_{xx},\qquad z=\eta.
\label{eq:icx}
\end{equation}
However, we may expand $\Phi(z=\eta)=\Phi(z=0)+(\partial \Phi/\partial z)_{z=0}\eta+O(\eta^2)$.  Because of the small-amplitude approximation, we can replace $\Phi(z=\eta)$ with $\Phi(z=0)$, and similarly for derivatives, giving 
\begin{equation}
\rho\frac{\partial\Phi}{\partial t}+\rho g\eta=\surften\eta_{xx},\qquad z=0.
\label{eq:icx1}
\end{equation}
The difference between Equations~\eqref{eq:icx} and~\eqref{eq:icx1} is subtle but it enables a great simplification in the foregoing analysis.

We now make the standard transformations:
\begin{subequations}
\begin{eqnarray}
\Phi&=&\Re\left[\phi(\vecx)\mathe^{-\imag\omega t}\right],\\
\eta&=&\Re\left[\widehat{\eta}(x)\mathe^{-\imag\omega t}\right].
\end{eqnarray}%
\label{eq:ansatzomega}%
\end{subequations}%
We henceforth drop the hat on $\widehat{\eta}$.  Thus, we use the same symbol for $\eta$ (which depends on $x$ and $t$), and $\widehat{\eta}$ (which depends on $x$ only). It should be clear from context which variable is being used.  In this way, Equation~\eqref{eq:icx1} becomes:
\begin{equation}
\rho\imag\omega\phi=\rho g\eta-\frac{\surften}{\rho}\eta_{xx},\qquad z=0.
\label{eq:ic1}
\end{equation}
A second interfacial condition is the kinematic condition.  In the small-amplitude approximation, which states that the interface moves with the flow, hence:
\begin{equation}
\frac{\partial\eta}{\partial t}+u\frac{\partial \eta}{\partial x}=w,\qquad z=\eta.
\end{equation}
As with Equation~\eqref{eq:icx1}, we linearize this identity on to the surface $z=0$, which gives:
\begin{equation}
\frac{\partial\eta}{\partial t}=w\qquad z=0,
\end{equation}
or $\partial_t \eta=\partial_z\phi$ on $z=0$, hence:
\begin{equation}
-\imag\omega\eta=\frac{\partial\phi}{\partial z},\qquad z=0.
\label{eq:ic2}
\end{equation}
We combine Equations~\eqref{eq:ic1}--\eqref{eq:ic2}.  First, Equation~\eqref{eq:ic2} gives $\eta=-1/(\imag\omega)\phi_z$.  We substitute this into Equation~\eqref{eq:ic1} to obtain a single boundary condition at $z=0$:
\begin{equation}
\omega^2\phi=g\frac{\partial\phi}{\partial z}-\frac{\surften}{\rho}\partial_{xx}\frac{\partial\phi}{\partial z},\qquad z=0.
\label{eq:ic_combo}
\end{equation}

\subsection{Solving Laplace's Equation}

We solve $\nabla^2\phi=0$ in the linearized domain $\omegalin=\{(x,z)| -h<z<0\}$.  We do separation of variables to get $\phi(x,z)=X(x)Z(z)$.  Following standard steps, we get:
\begin{equation}
\frac{X''}{X}=-\frac{Z''}{Z}=k^2.
\label{eq:sov}
\end{equation}
We look at the boundary conditions at $z=0$ next.  The boundary condition~\eqref{eq:ic_combo} gives:
\begin{equation}
\omega^2 X(x)Z(0)=\left(gX(x)-\frac{\surften}{\rho}X''(x)\right)Z'(0).
\end{equation}
We use the separation-of-variables condition~\eqref{eq:sov} to reduce this to:
\begin{equation}
\omega^2 Z(0)=\left(g-\frac{\surften}{\rho}k^2\right)Z'(0).
\end{equation}
We further re-write this as:
\begin{equation}
Z'(0)=\alpha_k Z(0),\qquad \alpha_k=\frac{\omega^2}{g-\frac{\surften}{\rho}k^2}.
\end{equation}
Putting it all together, we have to solve:
\begin{subequations}
\begin{eqnarray}
Z''+k^2Z&=&0,\\
Z'(-h)&=&0,\\
Z'(0)&=&\alpha_k Z(0).
\end{eqnarray}%
\end{subequations}
The solution is:
\begin{equation}
Z=\frac{\cos[k(z+h)]}{\cos kh},
\end{equation}
with solvability condition $k\tan(kh)=-\alpha_k$, or:
\begin{equation}
k\tan(kh)=-\frac{\omega^2}{g-\frac{\surften}{\rho}k^2}.
\label{eq:DR0}
\end{equation}
We label the solutions of Equation~\eqref{eq:DR1} as $k_n$, where $n\in\{0,1,2,\cdots\}$.

\subsection{Dispersion Relation}

\begin{figure}
	\centering
		\includegraphics[width=0.6\textwidth]{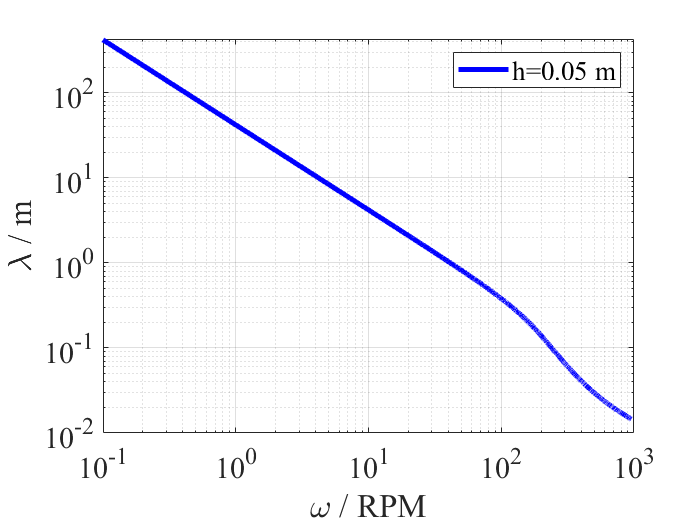}
		\caption{The dispersion relation~\eqref{eq:DR1}.  For a given $\omega$, there is a uniquely determined $k$-value, hence a uniquely determined wavelength $\lambda=2\pi/k$.  Parameter values: $h=0.05\,\mathrm{m}$, $\rho=1000,\,\mathrm{kg}\cdot\mathrm{m}^{-3}$, $g=9.8\,\mathrm{m}\cdot\mathrm{s}^{-2}$, $\surften=0.072\,\mathrm{N}\cdot\mathrm{m}^{-1}$.}
	\label{fig:DR_sample}
\end{figure}

Equation~\eqref{eq:DR0} has two solution types:
\begin{itemize}
\item Case 1.  This corresponds to $n=0$, so we are dealing with $k_0$.  In this case, $k_0$ is purely imaginary, and we write $k_0=\pm \imag \kappa$, where $\kappa$ is real.  Using the properties of trigonometric functions, Equation~\eqref{eq:DR1} reduces to:
\begin{equation}
\kappa\tanh(\kappa h)=\frac{\omega^2}{g+\frac{\surften}{\rho}\kappa^2},
\label{eq:DR1}
\end{equation}
which is precisely Equation~\eqref{eq:disp1}.  In this case, however, $\omega$ is known, and $\kappa$ has to be obtained by inversion.  A sample dispersion curve is shown in Figure~\ref{fig:DR_sample}.

\item Case 2.  In this case, we look at $k_n$, where $n\geq 1$. A standard graphical eigenvalue analysis shows in this case there are infinitely many real positive roots, confirming that $n\in\{1,2,\cdots\}$. 
\end{itemize}
Putting the two cases together, we have the following set of eigenfunctions, with $Z(z)$ being replaced by $\chi_n(z)$:
\begin{equation}
\chi_n(z)=\begin{cases} \frac{\cos [k_n(z+h)]}{\cos k_n h},&n\geq 1\\
                        \frac{\cosh [\kappa (z+h)]}{\cosh \kappa h},& n=0.\end{cases}
\label{eq:eigfuncs}
\end{equation}
As these are eigenfunctions of a self-adjoint operator, we have an orthogonality relation
\begin{equation}
\int_{-h}^0 \chi_m(z)\chi_n(z)\mathd z=C_n\delta_{nm}.
\end{equation}
In particular,
\begin{equation}
C_0=\frac{1}{4\kappa}\frac{1}{\cosh^2(\kappa h)}\left[2\kappa h+\sinh(2\kappa h)\right].
\label{eq:C0}
\end{equation}

\subsection{General Solution}

The general solution for the velocity potential can now be written as:
\begin{equation}
\phi(x,z)=\sum_{n=1}^\infty a_n \chi_n(z)\mathe^{-k_n x}+a_0\chi_0(z)\mathe^{\imag \kappa x}.
\end{equation}
Notice that we do not allow for a contribution proportional to $\mathe^{-\imag \kappa x}$, as this would correspond to a wave travelling inward from positive infinity, which is not physical. Furthermore, for a bounded solution, we rule out contributions that depend on $\mathe^{k_n x}$.   Thus, the Sommerfeld Radiation condition $\partial \phi/\partial x \sim \imag k \phi$ is satisfied as $x\rightarrow \infty$.    Furthermore, at $x=0$, we have:
\begin{equation}
\left(\frac{\partial\phi}{\partial x}\right)_{(x=0,z)}=\sum_{n=1}^\infty a_n \chi_n(z)(-k_n)+a_0\chi_0(z)(\imag \kappa).
\end{equation}
The boundary condition at $x=0$ is $\partial_x\phi=u=\partial_t\xi$, where $\xi$ is the displacement of the wall at $x=0$ (\textit{cf}. Equation~\eqref{eq:piston}).  Thus, we obtain:
\begin{equation}
\sum_{n=1}^\infty a_n \chi_n(z)(-k_n)+a_0\chi_0(z)(\imag \kappa)=f(z).
\end{equation}
Hence, the coefficients $a_0$ and $a_n$ can be determined from:
\begin{eqnarray*}
a_0&=&\frac{1}{(\imag \kappa)C_0}\int_{-h}^0 f(z)\chi_0(z)\mathd z,\\
a_n&=&\frac{1}{(-k_n) C_n}\int_{-h}^0 f(z)\chi_n(z)\mathd z,\qquad n\geq 1.
\end{eqnarray*}
In particular, for a piston wavemaker with $f(z)=f_0=\text{Const.}$, we have:
\begin{equation}
a_0=\frac{f_0}{(\imag \kappa)C_0} \frac{1}{\kappa}\frac{\sinh(\kappa h)}{\cosh(\kappa h)}.
\end{equation}
Furthermore, in the far field, we have
\begin{equation}
\phi\sim a_0 \chi_0(z)\mathe^{\imag \kappa x},\qquad x\rightarrow\infty,
\end{equation}
since $\mathe^{-k_nx}\rightarrow 0$ as $x\rightarrow \infty$, for $n\geq 1$.  
Only the oscillatory wave with dispersion relation~\eqref{eq:DR1} survives far downstream of the disturbance.

\subsection{Results of summary calculations}

By analysing the dispersion relation~\eqref{eq:DR1}, we can see what type of wavelengths can be expected for a given forcing frequency.  The wavelengths depend  on depth, as shown in Table~\ref{tab:DR}.
\begin{table}[htb]
	\centering
		%
		\begin{tabular}{|c|c|c|}
		\hline
		$\omega$ (RPM) & $\lambda$ ($h=0.05\,\mathrm{m}$) & $\lambda$ ($h=0.1\,\mathrm{m}$) \\
		\hline
		\hline
		10  & 4.19 & 5.91\\
		50  & 0.820 & 1.13\\
		100 & 0.381 & 0.484 \\
		200 & 0.139 & 0.142 \\
		\hline
		\end{tabular}
		\caption{Expected wavelengths (in metres), based on the dispersion relation~\eqref{eq:DR1}.  Depths: $h=0.05\,\mathrm{m}$ and $0.1\,\mathrm{m}$.  Other parameters as in 
		Figure~\ref{fig:DR_sample}.}
\label{tab:DR}
\end{table}

A further key quantity of interest is the height-to-stroke ratio, which we derive now for the piston wavemaker as follows.  We apply the kinematic condition~\eqref{eq:ic2} in the far field (for $x\rightarrow \infty$) to get
\begin{equation}
a_0 \left(\frac{\partial\chi_0}{\partial z}\right)_{z=0}=-\imag \omega \eta_0.
\end{equation}
Here, we have decomposed $\eta(x)$ into a phase $\eta_0$ and the complex exponential $\mathe^{\imag \kappa x}$, corresponding to the $n=0$ normal mode.
We fill in for $\chi_0(z)$ (\textit{cf.} Equation~\eqref{eq:eigfuncs}) to get
\begin{equation}
\frac{f_0}{(\imag \kappa) C_0} \frac{\sinh^2 (\kappa h)}{\cosh^2(\kappa h)}=-\imag \omega \eta_0.
\end{equation}
For a piston wavemaker, we have $f_0=\omega A \mathe^{\imag \varphi}$, where $A$ is the amplitude of the back-and-forth motion of the piston (and equal to half the stroke, $2A=S$), and $\varphi$ is a constant phase.  This gives:
\begin{equation}
\left|\frac{\eta}{A}\right|=\frac{1}{\kappa C_0}\frac{\sinh^2 (\kappa h)}{\cosh^2(\kappa h)},
\end{equation}
and filling in for $C_0$ gives:
\begin{equation}
\left|\frac{\eta_0}{A}\right|=\frac{4\sinh^2(\kappa h)}{2\kappa h+\sinh(2\kappa h)}.
\end{equation}
We identify the height of the wave $H=2|\eta_0|$, hence $|\eta/A|=|2\eta/(2A)|=H/S$.  This gives the required height-to-stroke ratio in the far field, valid for a piston wavemaker:
\begin{equation}
\frac{H}{S}=\frac{4\sinh^2(\kappa h)}{2\kappa h+\sinh(2\kappa h)}.
\label{eq:HoS}
\end{equation}

\section{Spatio-temporal analysis of small-amplitude water waves: the closed tank}
\label{sec:theory_closed}

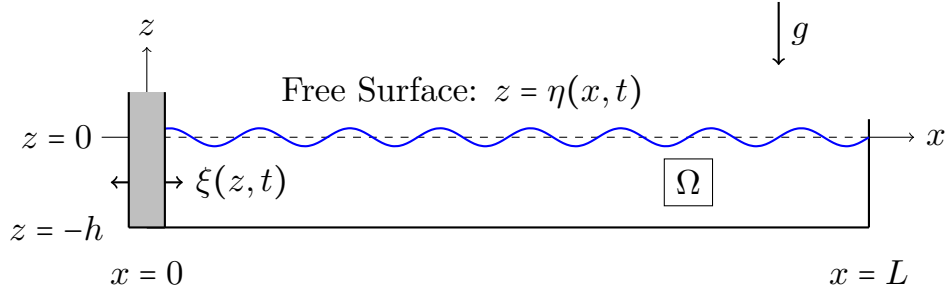
\begin{figure}
\centering
\begin{tikzpicture}[scale=1.2,transform shape]
    \draw[->] (0,0) -- (0,2) node[above] {$z$};

    \draw[thick] (0,0) -- (8,0); 
    \draw[thick] (8,0) -- (8,1.2);  
    
	\draw[-] (-0.5,1) -- (0,1);
    \draw[-,dashed] (0,1) -- (8,1); 
	\draw[->] (8,1)--(8.5,1);
	\node[right] at (8.5,1) {$x$};

    \draw[thick,blue,smooth,samples=100,domain=0.2:8] 
        plot(\x,{1+0.1*sin(2*pi*\x r)});
    
    \fill[gray!50] (-0.2,0) rectangle (0.2,1.5);
    \draw[thick] (-0.2,0) -- (-0.2,1.5);
    \draw[thick] (0.2,0) -- (0.2,1.5);
	\draw[thick] (-0.2,0) -- (0.2,0);
    
    \draw[->,thick] (0.2,0.5) -- (0.4,0.5) node[right] {$\xi(z,t)$};
    \draw[->,thick] (-0.2,0.5) -- (-0.4,0.5);

    \draw[->,thick] (7,2.5) -- (7,1.8) node[midway,right] {$g$};

    \node at (3.5,1.5) {Free Surface: $z=\eta(x,t)$};
    \node at (0,-0.5) {$x=0$};
	  \node at (8,-0.5) {$x=L$};
		
	\node at (-1,1) {$z=0$};
	\node at (-1,0) {$z=-h$};
	
	\node at (6,0.5) [rectangle,draw] {$\Omega$};
\end{tikzpicture}
\caption{Schematic diagram showing the generation of small-amplitude water waves by a piston wavemaker located at $x=0$ (closed tank)}
\label{fig:schematic2}
\end{figure}

In this section, we introduce wavemaker theory for a closed tank, in which $x\in [0,L]$, as shown schematically in Figure~\ref{fig:schematic2}.  The notation is the same as Section~\ref{sec:theory}.  The only difference is that a no-penetration boundary condition now applies at $x=L$:
\begin{equation}
\frac{\partial\phi}{\partial x}=0,\qquad x=L.
\label{eq:bc1}
\end{equation}
The left-hand boundary condition is unchanged from Section~\ref{sec:theory}, namely $\partial_x\phi=\partial_t\xi$, at $x=0$.  For a piston wavemaker, this amounts to:
%
%
\begin{equation}
\frac{\partial\phi}{\partial x}=\xiL,\qquad x=0.
\label{eq:bc2}
\end{equation}
The other boundary conditions are unchanged from before.  

\subsection{Cosine Transform}

We take Laplace's Equation $\nabla^2\phi=0$ in the linearized domain $\omegalin=\{(x,z)|-h<z<0,0<x<L\}$, multiply by $\cos(n\pi x/L)$ and integrate with respect to $x$ from $x=0$ to $x=L$.  Here, $n\in \{0,1,\cdots\}$.  Applying the boundary conditions~\eqref{eq:bc1} and~\eqref{eq:bc2}, we obtain:
\begin{equation}
\frac{\mathd^2\phihatn}{\mathd z^2}-k_n^2\phihatn=\xiL.
\end{equation}
We solve this equation subject to the no-penetration boundary condition $\mathd\phihatn/\mathd z=0$ at $z=-h$ and the dynamic boundary condition~\eqref{eq:ic_combo} at $z=0$.  This gives:
\begin{itemize}[noitemsep]
\item The case $n=0$:
\begin{equation}
\phihatzero=\frac{g\xiL h}{\omega^2}+\xiL h z+\tfrac{1}{2}\xiL z^2,\qquad \omega\neq 0.
\label{eq:n0sln}
\end{equation}
\item The case $n>0$:
\begin{equation}
\phihatn=A_n\cosh[k_n(z+h)]-\frac{\xiL}{k_n^2},
\end{equation}
where
\begin{equation}
A_n= \frac{ \omega^2 \xiL }{k_n^2\cosh(k_n h)\bigg\{ \omega^2- \left[g+(\gamma/\rho)k_n^2\right]k_n\tanh k_n \bigg\}}.
\label{eq:n_sln}
\end{equation}
\end{itemize}
Equations~\eqref{eq:n0sln}--\eqref{eq:n_sln} are valid off-resonance, that is, provided 
\begin{equation}
 \left[g+(\gamma/\rho)k_n^2\right]k_n\tanh k_n \neq \omega^2,
\end{equation}
and provided also that $\omega \neq 0$.  On resonance, the original trial solution~\eqref{eq:ansatzomega} which sets $\Phi=\Re\left[\phi(\vecx)\mathe^{-\imag\omega t}\right]$ 
and $\eta=\Re\left[\widehat{\eta}(x)\mathe^{-\imag\omega t}\right]$, is no longer valid.  In this case, the solution needs to be derived by carefully following the steps in a Laplace-transform calculation~\cite{lee1989transient}.  This yields a resonant solution which grows algebraically in time.  


\subsection{General Solution}

We use the inverse cosine transformation to write the general solution for $\phi(x,z)$:
\begin{multline}
\phi(x,z)=\frac{1}{L}\left[ \frac{g\xiL h}{\omega^2}+\xiL h z+\tfrac{1}{2}\xiL z^2\right]\\+
\frac{2}{L}\sum_{n=1}^\infty \frac{ \omega^2 \xiL }{k_n^2\cosh(k_n h)\bigg\{ \omega^2-\left[g+(\gamma/\rho)k_n^2\right]k_n\tanh k_n \bigg\}}\cosh[k_n(z+h)]\cos(n\pi x/L).
\end{multline}
The free surface  is given by $\eta=[-1/(\imag\omega)\phi_z]_{z=0}$, hence:
\begin{multline}
\eta(x,t)=\Re\left[ \frac{\imag \xiL}{\omega} \mathe^{-\imag\omega t}\right]\\+
\frac{2}{L}\sum_{n=1}^\infty \frac{ \Re(\imag \xiL \omega \mathe^{-\imag\omega t} )}{k_n\cosh(k_n h)\bigg\{ \omega^2-\left[g+(\gamma/\rho)k_n^2\right]k_n\tanh k_n \bigg\}}\sinh(k_n h)\cos(n\pi x/L).
\label{eq:etax}
\end{multline}

\subsection{Approximate Solution}

\begin{figure}[htb]
	\centering
		\includegraphics[width=0.7\textwidth]{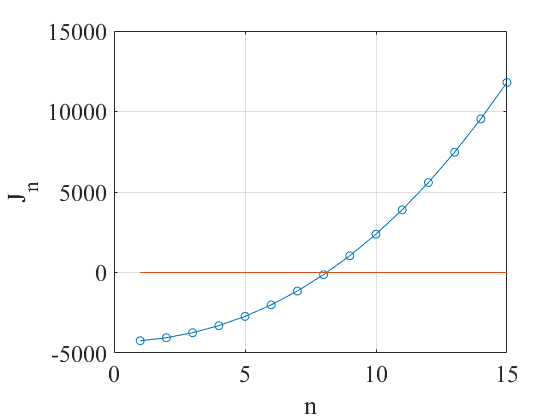}
		\caption{Plot showing the minimization of $J_n$ at $n=8$ for the case $\gamma=0.072\,\mathrm{N}\cdot\mathrm{m}^{-1}$, $\rho=1000\,\mathrm{kg}\cdot\mathrm{m}^{-3}$, $g=9.8\,\mathrm{m}\cdot\mathrm{s}^{-2}$, $L=1\,\mathrm{m}$, $h=0.05\,\mathrm{m}$, and $\omega=140\,\mathrm{RPM}$.}
	\label{fig:Jn}
\end{figure}

In this section, we introduce an approximate solution of Equation~\eqref{eq:etax} which contains only one wavenumber.  This is obtained simply by taking the most-dominant component of the sum in Equation~\eqref{eq:etax}:
\begin{multline}
\eta(x,t)=\Re\left[ \frac{\imag \xiL}{\omega} \mathe^{-\imag\omega t}\right]\\+
\frac{2}{L} \frac{ \Re(\imag \xiL \omega \mathe^{-\imag\omega t} )}{k_n\cosh(k_{n_0} h)\bigg\{ \omega^2-\left[g+(\gamma/\rho)k_{n_0}^2\right]k_{n_0}\tanh k_{n_0} \bigg\}}\sinh(k_{n_0} h)\cos(n_0\pi x/L),
\label{eq:etax2}
\end{multline}
where $n_0$ is the solution of
\begin{multline}
n_0=\mathrm{arg}\min_{n\in\mathbb{N}}J_n,\\
\phantom{aaaaaaaa}\text{where }J_n=k_n[\cosh(k_n h)/\sinh(k_n h)]\bigg\{ \left[g+(\gamma/\rho)k_n^2\right]k_n\tanh k_n - \omega^2\bigg\}.
\end{multline}
Referring to Figure~\ref{fig:Jn}, for $\gamma=0.072\,\mathrm{N}\cdot\mathrm{m}^{-1}$, $\rho=1000\,\mathrm{kg}\cdot\mathrm{m}^{-3}$, $g=9.8\,\mathrm{m}\cdot\mathrm{s}^{-2}$, $L=1\,\mathrm{m}$, $h=0.05\,\mathrm{m}$, and $\omega=140\,\mathrm{RPM}$, the integer that minimizes $J_n$ is $n_0=8$, corresponding to four maxima in the wave tank.
Furthermore, the neighbouring values of $J_{n_0\pm 1}$ are well separated from $J_{n_0}$, with $|J_{n_0\pm 1}/J_{n_0}|>7.6$, meaning that Equation~\eqref{eq:etax2} is a good first approximation.

\section{Experiments using the Tabletop Flume}
\label{sec:flume}

In this section we report on a study of waves generated using the tabletop flume described in the introduction.  We set out the technical specification of the flume and the associated wavemaker, and report summary results.  An in-depth statistical analysis is subsequently carried out, which produces excellent agreement with between the theory in Sections~\ref{sec:theory}--\ref{sec:theory_closed} and the measurements.

\subsection{Specification of the Flume and the Wavemaker}

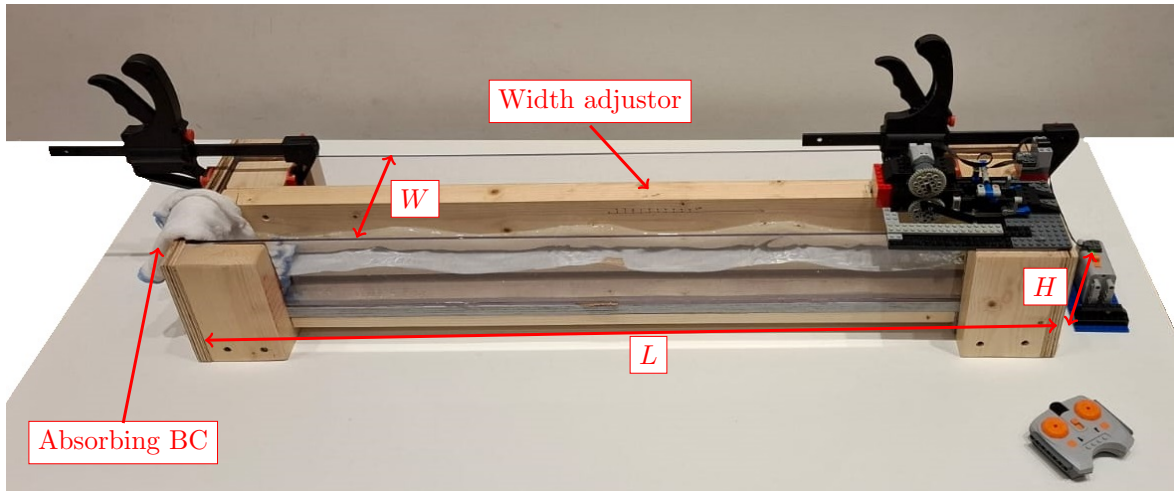
\begin{figure}[htb]
\centering
\resizebox{0.95\textwidth}{!}{%
\begin{tikzpicture}
    \node[anchor=south west, inner sep=0] (background) at (0,0) {
        \includegraphics[width=\textwidth]{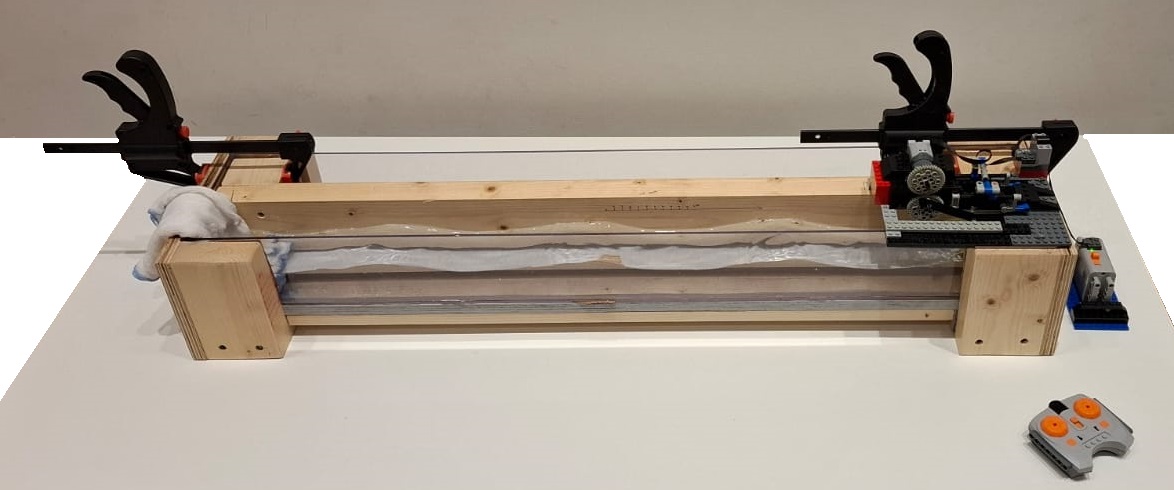}
    };
    
    \begin{scope}[x={(background.south east)},y={(background.north west)}]
				\node at (0.35,0.6) [rectangle,draw,red,fill=white]  {$W$};
				\draw[<->,red,very thick] (0.3,0.52) -- (0.33,0.69);
				\node at (0.55,0.28) [rectangle,draw,red,fill=white]  {$L$};
				\draw[<->,red,very thick] (0.17,0.31) -- (0.9,0.34);
			  \node at (0.89,0.42) [rectangle,draw,red,fill=white]  {$H$};
				\draw[<->,red,very thick] (0.91,0.34) -- (0.93,0.49);
				\node at (0.5,0.8) [rectangle,draw,red,fill=white]  {Width adjustor};
				\draw[->,red,very thick] (0.5,0.75) -- (0.55,0.62);
				\node at (0.1,0.1) [rectangle,draw,red,fill=white]  {Absorbing BC};
				\draw[->,red,very thick] (0.1,0.14) -- (0.13,0.5);
    \end{scope}
\end{tikzpicture}
} 
\caption{Photograph of the tabletop flume with wavemaker attached}
\label{fig:wt1}
\end{figure}

\begin{figure}
	\centering
		\includegraphics[width=0.7\textwidth]{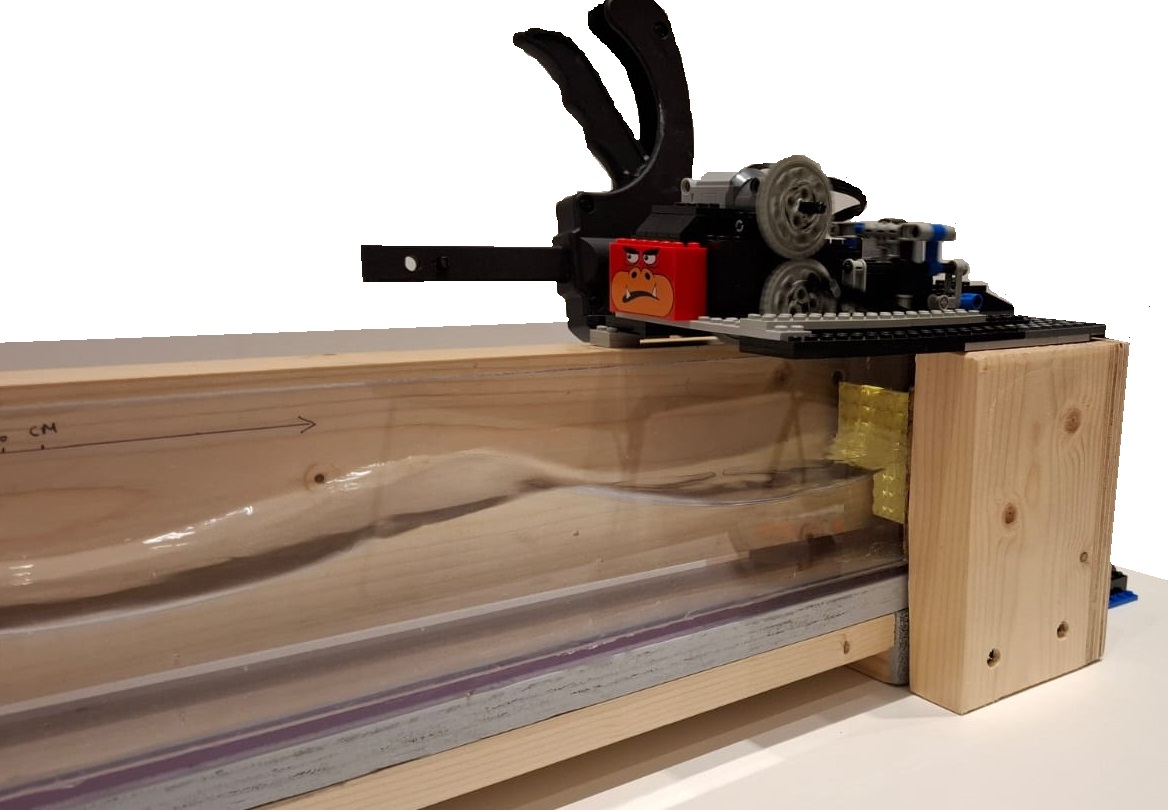}
		\caption{A zoom-in on Figure~\ref{fig:wt1} showing the back-and-forth action of the flap (yellow Lego board) which generates the wave motion.  }
	\label{fig:wt2}
\end{figure}

The tabletop flume is shown in Figure~\ref{fig:wt1} and consists of a perspex box (open at the top) of length $L=1\,\mathrm{m}$, height $H=0.1\,\mathrm{m}$, and width $W=0.16\,\mathrm{m}$.  These measurements are taken with respect to the outer edges of the box.  As the perspex has a thickness $4\,\mathrm{mm}$, the inner dimensions of the box are slightly smaller.     For added strength, the perspex box is set in a wooden housing.     To produce two-dimensional waves, the width of the flume can reduced by way of a simple width-adjuster made up of a wooden beam clamped at both ends to the flume housing.  To help with measurements and analysis, rulers on the $\mathrm{cm}$ scale have been drawn one on the flume housing (for water depth) and one on the width adjustor (for wavelength) with permanent marker.  This enables us to do quantitative video analysis at a later stage.   A photograph of the setup is shown in Figure~\ref{fig:wt1}.
A piston wave-maker is secured to the flume housing as shown Figure~\ref{fig:wt1}.  The back-and-forth action of the flap which generates the wave motion is shown in Figure~\ref{fig:wt2}.  
 At the opposite end of the flume, an absorbing boundary condition is applied by the simple expedient of draping a cloth over the end, as shown also in the figure.

\begin{figure}[htb]
\centering
\resizebox{0.7\textwidth}{!}{%
\begin{tikzpicture}
    \node[anchor=south west, inner sep=0] (background) at (0,0) {
        \includegraphics[width=\textwidth]{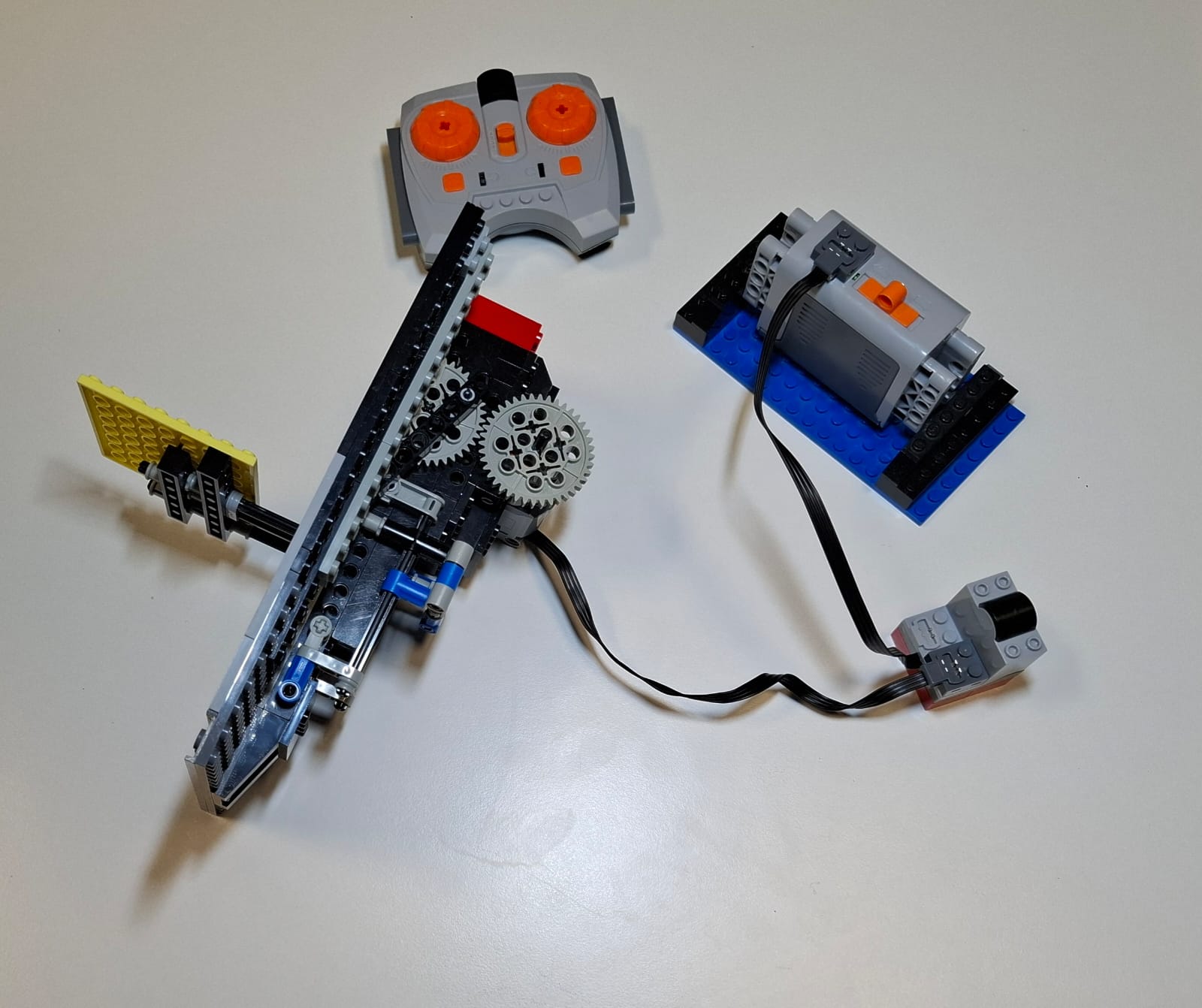}
    };
    
    \begin{scope}[x={(background.south east)},y={(background.north west)}]
				\node at (0.3,0.95) [rectangle,draw,red,fill=white]  {(3a)};
				\node at (0.85,0.7) [rectangle,draw,red,fill=white]  {(2)};
				\node at (0.9,0.4) [rectangle,draw,red,fill=white]  {(3b)};
				\node at (0.5,0.62) [rectangle,draw,red,fill=white]  {(4)};
				\node at (0.43,0.4) [rectangle,draw,red,fill=white]  {(1)};
				\draw[->,red,very thick] (0.43,0.425) -- (0.435,0.47);
				\node at (0.2,0.6) [rectangle,draw,red,fill=white]  {(5)};
				\node at (0.2,0.15) [rectangle,draw,red,fill=white]  {(6)};
    \end{scope}
\end{tikzpicture}
} 
\caption{Photograph showing the different components Lego wavemaker}
\label{fig:wt3}
\end{figure}
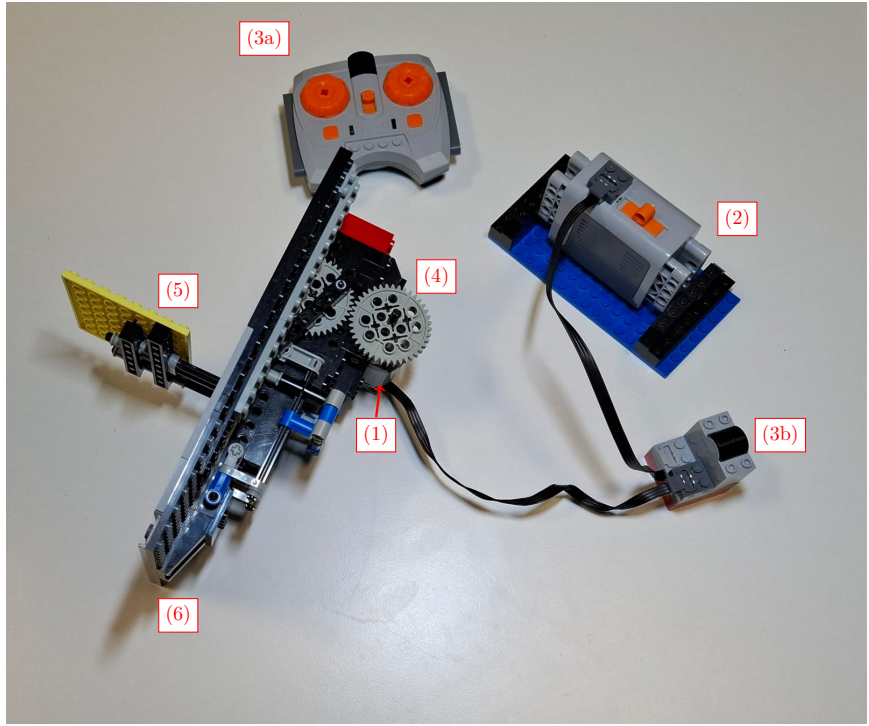

The wavemaker is a piston wavemaker made from Lego Tecnic components.  Detailed instructions to make the wave-maker are provided in the accompanying GitHub repository~\cite{waveFlume}.  The wave-maker consists of five parts, highlighted as follows and shown separately in Figure~\ref{fig:wt3}:
\begin{enumerate}[noitemsep]
\item Electric Motor;
\item Battery pack;
\item Variable RPM controller (parts (3a) and (3b) in  Figure~\ref{fig:wt2});
\item Transmission system, with $1:1$ gear ratio;
\item Oscillating piston;
\item Housing.
\end{enumerate}
This Lego wavemaker is screwed to the wooden housing which encases the perspex box.

\subsection{Preliminary Results}

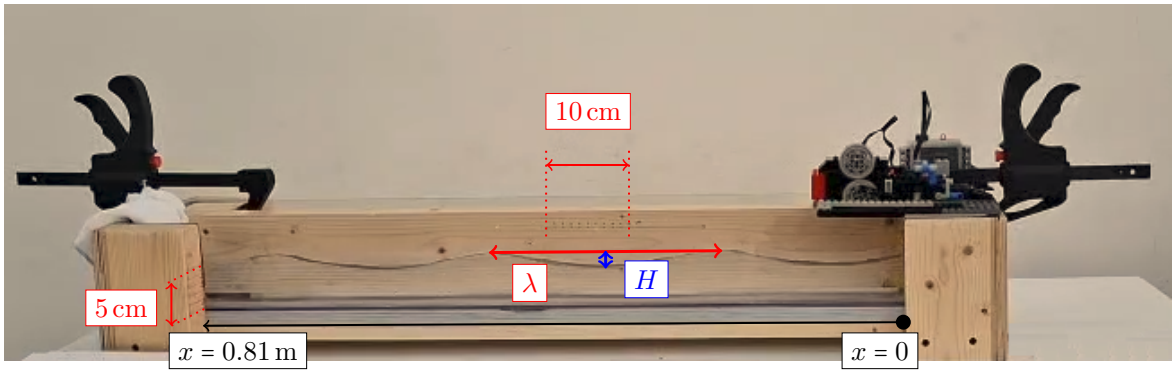
\begin{figure}[htb]
\centering
\resizebox{0.95\textwidth}{!}{%
\begin{tikzpicture}
    \node[anchor=south west, inner sep=0] (background) at (0,0) {
        \includegraphics[width=\textwidth]{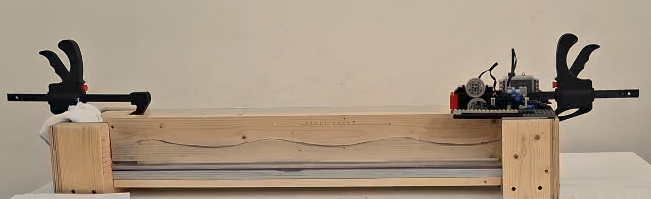}
    };
    
    \begin{scope}[x={(background.south east)},y={(background.north west)}]
	      %
				\draw[<->,red,very thick] (0.415,0.305) -- (0.615,0.31);
				\node at (0.45,0.22) [rectangle,draw,red,fill=white]  {$\lambda$};
				\draw[<->,blue,very thick] (0.515,0.26) -- (0.515,0.31);
				\node at (0.55,0.23) [rectangle,draw,blue,fill=white]  {$H$};
				\draw[dotted,red,thick] (0.464,0.35) -- (0.464,0.6);
				\draw[dotted,red,thick] (0.535,0.35) -- (0.535,0.6);
				\draw[<->,red,thick] (0.464,0.55) -- (0.535,0.55);
				\node at (0.5,0.7) [rectangle,draw,red,fill=white]  {$10\,\mathrm{cm}$};
				\draw[dotted,red,thick] (0.12+0.023,0.1+0.0) -- (0.15+0.023,0.15+0.0);
				\draw[dotted,red,thick] (0.12+0.02,0.1+0.12) -- (0.15+0.02,0.15+0.12);
				\draw[<->,red,thick] (0.12+0.023,0.1+0.0) -- (0.12+0.023,0.1+0.12);
				\node at (0.1,0.15) [rectangle,draw,red,fill=white]  {$5\,\mathrm{cm}$};
				\draw[->,black,thick] (0.77,0.11) -- (0.17,0.1);
				\fill (0.77,0.11) circle (3pt); 
				%
				%
				\node at (0.75,0.03) [rectangle,draw,fill=white]{$x=0$};
				\node at (0.2,0.03) [rectangle,draw,fill=white]{$x=0.81\,\mathrm{m}$};
    \end{scope}
\end{tikzpicture}
} 
\caption{Snapshot of the wavemaker experiment, Frame 7 from Reference~\cite{onaraighWave}.  A re-centred $x$-axis is shown; the location of $x=0$ is at the black dot.}
\label{fig:measure_k}
\end{figure}

We present here a first set of results  based on  the setup shown in Figure~\ref{fig:wt1}.  The water depth is set as $h=5.0\,\mathrm{cm}$, correct to the nearest millimetre.  Hence, to avoid spurious precision, we report final results here to two significant figures only.  Intermediate measurements may be reported to higher precision, if available.
A video recording was made of the experiment using a Samsung Galaxy A55.   To enable further analysis and reproducibility, the video has been posted on YouTube~\cite{onaraighWave}.  The recording was analyzed on a frame-by-frame basis using the \texttt{VideoReader} function in Matlab.  From this analysis, the frame rate of video is obtained:  $r=29.86\,\mathrm{fps}$.
  From a representative frame (Frame 7 in the video, see Figure~\ref{fig:measure_k}), the  wavelength is measured to be $0.28\pm 0.01\,\mathrm{m}$.  Correspondingly, $k=(2\pi/\lambda)\pm \Delta k$, where $\Delta k=k(\Delta \lambda/\lambda)$, hence $k=(22\pm 1)\,\mathrm{m}^{-1}$.
	
Although the setup shown in Figures~\ref{fig:wt1}--\ref{fig:wt3} allows for variable RPM of the piston wavemaker, given the load on the wavemaker, the RPM is not known \textit{a priori}.  For this reason, we have also measure the frequency of the wavemaker  from the video analysis: one cycle of the piston is measured to take $13\pm 1$ frames,  the measurement error here coming from the uncertainty in identifying which is the last frame in one cycle of the piston.   These data then give the period of the piston wavemaker as $T=\left[(13\pm 1)/r\right]\,\mathrm{s}$, where $\Delta T$ is identified here as $(1/r)\,\mathrm{s}$. The frequency is therefore calculated as:
\begin{equation}
\omega=\frac{2\pi}{T}\left(1\pm\frac{\Delta T}{T}\right)\,\mathrm{rad}\cdot\mathrm{s}^{-1},
\end{equation}
hence $\omega=140\pm 10\,\mathrm{RPM}$.  

We jump the gun slightly and endeavour to connect these measurements back to the theory of \textit{travelling waves} in Section~\ref{sec:theory} (full justification below -- in the in-depth statistical analysis).
For this purpose, we  refer the reader to Table~\ref{tab:DR1}.  From this table, we identify a range of wavelengths  $I_{theory}=[0.22,0.27]\,\mathrm{m}$ consistent with the travelling-wave theory.  Furthermore, we identity $I_{measured}=[0.27,0.29]\,\mathrm{m}$ as the range of wavelengths consistent with the experimental measurements.  Since $I_{theory}\cap I_{measured}$ is non-empty, the
 the experimental results are consistent with the theoretical analysis.
\begin{table}[htb]
	\centering
		\begin{tabular}{|c|c|c|}
		\hline
		$\lambda(\omega=130\,\mathrm{RPM})$ & $\lambda(\omega=140\,\mathrm{RPM})$ & $\lambda(\omega=150\,\mathrm{RPM})$ \\
		\hline
		\hline
		$0.27\,\mathrm{m}$ & $0.25\,\mathrm{m}$ & $0.22\,\mathrm{m}$\\
		\hline
		\end{tabular}
		\caption{Range of wavelengths consistent with the linear theory.  Water depth: $h=0.05\,\mathrm{m}$ and all other parameters the same as in Figure~\ref{fig:DR_sample}.}
\label{tab:DR1}
\end{table}

We have also measured the wave height as $H=(1.8\pm 0.3)\,\mathrm{cm}$ 
and the stroke length of the piston as $S=(2.4\pm 0.2)\,\mathrm{cm}$, giving $(H/S)_{\text{inferred}}=0.8\pm 0.2$.
From Equation~\eqref{eq:HoS} for travelling waves (with $\lambda=0.28\,\mathrm{m}$), we have $(H/S)_{\text{theory}}=1.09\pm 0.04$.  Thus, the measured value of the height-to-stroke ratio is not consistent with the theoretical value. 
To understand this discrepancy, we analyze our results in more detail in what follows.

\subsection{In-depth statistical analysis}

\begin{figure}
	\centering
		\includegraphics[width=0.8\textwidth]{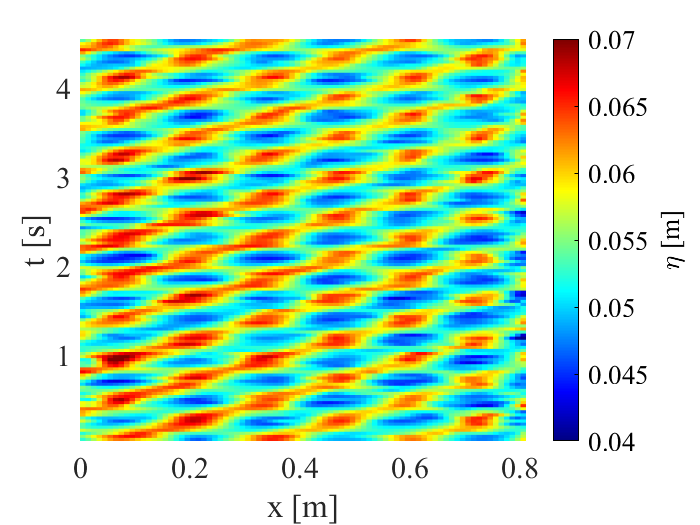}
		\caption{Plot of the free-surface height $\eta$ as a function of space and time}
	\label{fig:eta_spacetime_data}
\end{figure}

To understand the results in the  video in more detail, we have digitized the first 120 frames  and made a record of the interface height on a frame-by-frame basis, and stored the results in a space-time array $\eta(x,t)$.  Here, $x\in [0,0.81]$ is the coordinate along the horizontal direction in the test section shown in Figure~\ref{fig:measure_k}, and $t$ is time.  The resulting data have been made available in the accompanying GitHub repository~\cite{waveFlume}.   To avoid spurious precision, we again report the results to two significant figures only.  Intermediate parameter values may be reported to higher precision.  Data are stored at discrete spatial coordinates $x_i$ and discrete times $t_j$.  We plot the results of this digitization procedure in Figure~\ref{fig:eta_spacetime_data}.  

\begin{figure}
	\centering
		\includegraphics[width=0.8\textwidth]{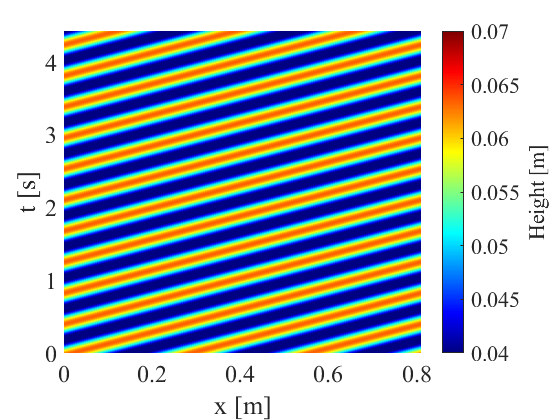}
		\caption{Plot of the model free-surface height $\eta_{model}$ (Equation~\eqref{eq:model1}) as a function of space and time}
	\label{fig:model1}
\end{figure}

From Figure~\ref{fig:eta_spacetime_data}, a well-defined train of travelling waves can be picked out.  Using non-linear least squares, we can fit a functional form
\begin{equation}
\eta_{model}(x,t)=h_0+A\cos(\omega t - kx+\varphi)
\label{eq:model1}
\end{equation}
to the data.   The non-linear least squares problem is solved by minimizing the cost function
\begin{equation}
J(h_0,A,\omega,\phi)=\sum_{i}\sum_{j} \left[\eta_{model}(x_i,t_j)-\eta(x_i,t_j)\right]^2.
\label{eq:cost1}
\end{equation}
The values that minimize the cost function are the `fitting parameters', these are given in Table~\ref{tab:fitting1}.  The upper and lower bounds are obtained by statistical bootstrapping, and correspond to the $2.5\%$ and $97.5\%$ confidence intervals generated using that method.  The Matlab files used for the bootstrapping are made available in the accompanying GitHub repository~\cite{waveFlume}.
\begin{table}
	\centering
		\begin{tabular}{|c|c|c|c|c|c|}
		\hline
		                  & $h_0\,\left[\mathrm{m}\right]$ &  $A\,\left[\mathrm{m}\right]$  & 
											$\omega\,\left[\mathrm{rad}\cdot\mathrm{s}^{-1}\right]$  & $k\,\left[\mathrm{m}^{-1}\right]$ & $\varphi$ \\
		\hline
		\hline
		Best Estimate     &0.0500    & 0.0134    &   14.7     & 23.3  & 0.000  \\
		Lower Bound       &0.0499    & 0.0083    &   14.5     & 23.3  & 0.510 \\
		Upper Bound       &0.0502    & 0.0168    &   14.9     & 25.4  & 0.819  \\
		\hline
		\end{tabular}
		\caption{Optimal fitting parameters, model~\eqref{eq:model1}}
\label{tab:fitting1}
\end{table}
A space-time plot of the model profile is shown in Figure~\ref{fig:model1}.
The fitted value of $k$ implies $\lambda=2\pi/k \in [0.25,0.27]\,\mathrm{m}=I_{measured}$, with a best estimate $\lambda=0.27\,\mathrm{m}$.  This is in the same range as our previous summary measurement of the wavelength in Figure~\ref{fig:measure_k}.

Next, we calculate the theoretical wavelength for travelling waves,
evaluated at the best-estimate frequency $\omega=14.7\,\mathrm{rad}\cdot\mathrm{s}^{-1}=140\,\mathrm{RPM}$; this is $\lambda=0.25\,\mathrm{m}$.  Using the confidence intervals in Table~\ref{tab:fitting1}, we are able to  account for the spread in the estimated value of $\omega$.  These are inputted into the travelling-wave theory to produce an allowed interval of wavelengths $I_{theory}= [0.24,0.25]\,\mathrm{m}$ (Table~\ref{tab:DR1_updated}).  Since the overlap $I_{theory}\cap I_{measured}$ is non-empty,  the travelling-wave theory is consistent with the measurements.

\begin{table}[htb]
	\centering
		\begin{tabular}{|c|c|c|}
		\hline
		$\lambda(\omega=138\,\mathrm{RPM})$ & $\lambda(\omega=140\,\mathrm{RPM})$ & $\lambda(\omega=142\,\mathrm{RPM})$ \\
		\hline
		\hline
		$0.25\,\mathrm{m}$ & $0.25\,\mathrm{m}$ & $0.24\,\mathrm{m}$\\
		\hline
		\end{tabular}
		\caption{Range of wavelengths consistent with the linear theory (updated, using the model~\eqref{eq:model1}).   Water depth: $h=0.050\,\mathrm{m}$ and all other parameters the same as in Figure~\ref{fig:DR_sample}.   }
\label{tab:DR1_updated}
\end{table}

\begin{figure}
	\centering
		\includegraphics[width=0.8\textwidth]{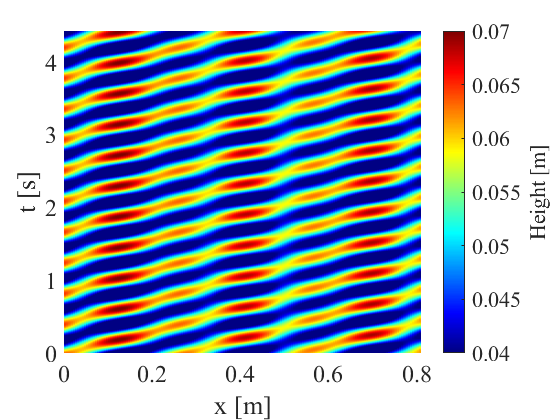}
		\caption{Plot of the model free-surface height $\eta_{model}$ (Equation~\eqref{eq:model2}) as a function of space and time}
	\label{fig:model2}
\end{figure}

Notwithstanding the accurate measurement of $\lambda$ obtained from fitting the model~\eqref{eq:model1} to the data, there is a clear visual mismatch between Figure~\ref{fig:eta_spacetime_data} (experiment) and~\ref{fig:model1} (model).  In the experimental data, there is clear evidence of a standing wave, in addition to the travelling wave.  Therefore, to obtain better agreement between the model and the data, we perform non-linear least squares again, and we fit a functional form
\begin{equation}
\eta_{model}(x,t)=h_0+A_1\cos(\omega t - k_1 x+\varphi_1)+A_2\cos(\omega t+\varphi_2)\cos(k_2 x)+A_3\cos(\omega t+\varphi_3)
\label{eq:model2}
\end{equation}
to the data.  The non-linear least squares problem is solved by minimizing a cost function analogous to Equation~\eqref{eq:cost1}.  The fitting parameters are given in 
Table~\ref{tab:fitting2}.   The upper and lower bounds are again obtained by statistical bootstrapping, and correspond to the $2.5\%$ and $97.5\%$ confidence intervals generated using that method.
\begin{table}
	\centering
		\begin{tabular}{|c|c|c|c|c|c|}
		\hline
		                  & $h_0\,\left[\mathrm{m}\right]$ &  $A_1\,\left[\mathrm{m}\right]$  & 
											$\omega\,\left[\mathrm{rad}\cdot\mathrm{s}^{-1}\right]$  & $k_1\,\left[\mathrm{m}^{-1}\right]$ & $\varphi_1$ \\
		\hline
		\hline
		Best Estimate     &0.0500    & 0.0129   &   14.7  & 24.6 & 0.152  \\
		Lower Bound       &0.0499    & 0.00531  &   14.4  & 23.4 & 0.000 \\
		Upper Bound       &0.0501    & 0.0154   &   14.9  & 26.5  & 0.750  \\
		\hline
		\end{tabular}
\[
\phantom{a}
\]

		\begin{tabular}{|c|c|c|c|c|c|}
		\hline
		                  &  $A_2\,\left[\mathrm{m}\right]$   
											& $k_2\,\left[\mathrm{m}^{-1}\right]$  
											& $\varphi_2$ 
											& $A_3\,\left[\mathrm{m}\right]$ 
											& $\varphi_3$ \\
		\hline
		\hline
		Best Estimate     & 0.00503 & 22.19   & 2.74  & 0.00308 & 3.36 \\ 
		Lower Bound       & 0.00304 & 15.7   & 1.37  & 0.00280 & 3.14\\ 
		Upper Bound       & 0.00617 & 28.1   & 5.22  & 0.00350 & 4.02\\ 
		\hline
		\end{tabular}
		\caption{Best-fit parameters, model~\eqref{eq:model2}.  The lower bounds are obtained from the $2.5\%$ and $97.5\%$ confidence intervals.  }
\label{tab:fitting2}
\end{table}
 Results are shown in Figure~\ref{fig:model2}.
There is much better qualitative agreement between Figure~\ref{fig:eta_spacetime_data} (experiment) and~\ref{fig:model2} (updated model), than there is between the experiment and the previous travelling-wave-only model.  Therefore, the experimental evidence is that the wave profile in the tank is a linear superposition of a travelling wave (obtained by waves traveling to the end of the tank and being absorbed by the cloth), and a standing wave (obtained by waves reflected back-and-forth between both ends of the tank).  


Referring to Table~\ref{tab:fitting2}, $\lambda=2\pi/k\in [0.24 ,0.27]\,\mathrm{m}=I_{measured}$, with a best estimate $\lambda=0.26\,\mathrm{m}$, consistent with the previous fitted model~\eqref{eq:model1}.  
%
%
%
Repeating earlier, similar calculations, we calculate the theoretical wavelength (travelling waves) evaluated at the best-estimate frequency $\omega=14.7\,\mathrm{rad}\cdot\mathrm{s}^{-1}=140\,\mathrm{RPM}$; this is $\lambda=0.25\,\mathrm{m}$. 
 We again account for the spread in the fitted value of $\omega$, using the $\omega$-confidence intervals in Table~\ref{tab:fitting2} and the results in Table~\ref{tab:DR1_updated}, hence $I_{theory}=[0.24,0.25]\,\mathrm{m}$.   Again, the overlap $I_{theory}\cap I_{measured}$ is non-empty, so the measurements are consistent with the theory.

%
%
%

We look at the height-to-stroke formula in the context of the fitted model~\eqref{eq:model2}.  The appropriate value of the  height is now the height of the travelling-wave component, hence $H=2A_1$.  
 We use $S=(2.4\pm 0.2)\,\mathrm{cm}$, as previously. 
Hence, the inferred height-to-stroke ratio is  $(H/S)_{\mathrm{inferred}}=1.07$.  However, allowing for the uncertainty in the fitted value of $A_1$ and the measured value of $S$, we have $(H/S)_{\mathrm{inferred}}\in [0.41,1.29]$.
 From Equation~\eqref{eq:HoS} with $\lambda=0.25\,\mathrm{m}$, we have 
$(H/S)_{\mathrm{theory}}=1.21$, so the measured and theoretical values are consistent.



\subsection{Discussion}

The observed waveform in Figure~\ref{fig:eta_spacetime_data} clearly consists of a superposition of a travelling wave and a standing wave.  The theory in Section~\ref{sec:theory} describes a travelling wave only and can be realised by having a perfectly absorbing boundary condition at the far end of the wave tank, absorbing all outgoing waves.  The opposite extreme is the theory in Section~\ref{sec:theory_closed}, which describes a standing wave.  This setup can be realised by having a perfectly reflecting boundary condition at the far end of the tank.

It is clear from the results in Figure~\ref{fig:eta_spacetime_data} that the cloth placed at the end of the wave tank does a good job of absorbing outgoing waves but is not perfect.  Therefore, the boundary condition at $x=L$ is not known \textit{a priori} but is such that the allowed waves in the tank are a superposition of standing and travelling waves.  Thus, the two complementary theories in Section~\ref{sec:theory} and~\ref{sec:theory_closed} are required to explain the observed waveform.  This is the inspiration for the fitted model in Equation~\eqref{eq:model2}.  

Referring back to the fitted model in Equation~\eqref{eq:model2}, the estimate for the standing-wave wavenumber $k_2$ is in the range $k_2\in [15.7,28.1]\,\mathrm{m}^{-1}$.  Referring to the standing-wave theory in Section~\ref{sec:theory_closed}, we identify this as $k_2=n\pi/L$.
%
%
%
This gives $n\in [5,9]$, which is consistent with the predicted value of the most-dominant standing-wave mode in Section~\ref{sec:theory_closed} ($n=8$, Figure~\ref{fig:Jn}).

In the next section we will look at a computational model for the wave tank which allows for much more precise control of the outgoing waves and hence, a much sharper travelling-wave solution in the (numerical) wave tank.





\section{Computational Fluid Dynamics using OpenFOAM}
\label{sec:cfd}

To understand the waves generated by the tabletop flume in more detail, we perform numerical simulations  using a two-phase flow algorithm in the open-source finite-volume code OpenFOAM.  In this section, we report on our findings.  We first give a broad description two-phase flow modelling to wave modelling, we describe the implementation of the model in OpenFOAM and then present our results.

\subsection{Volume-of-Fluid Method}

We describe here the Volume-of-Fluid (VoF) method to model two-phase flow.  The method is based on a one-fluid formulation of the two-phase Navier--Stokes equations, first introduced by Brackbill~\cite{brackbill1992continuum}.  In such  a one-fluid formulation, one solves a single set of Navier--Stokes equations for a single fluid, the properties of which transition sharply across a zone corresponding to the interface in the classic two-fluid formulation.  An advantage of this approach is that one does not have to solve separate Navier--Stokes equations in the different phases, which would be computationally challenging and would require the prescription of complicated matching conditions across the interface separating the phases.

  To build such a one-fluid formulation, one starts with an indicator function $\chi(\vecx)$ which tracks the phases:
\begin{equation}
\label{eq1}
\chi(\vecx) = 
\begin{cases}
    1 & \text{if } \vecx \text{ is in the liquid phase}, \\
    0 & \text{if } \vecx \text{ is in the gas phase}, \\
\end{cases}
\end{equation}
Since $\chi(\vecx)$ is a step function, it cannot be differentiated numerically.  Hence, we introduce a smoothened version:
\begin{equation}
\label{eq2}
\alpha(\vecx) = \frac{1}{V} \int_V \chi(\vecx') \, d^3\vecx',
\end{equation}
where $V$ is a small test volume.  Thus, in the Volume-of-Fluid formulation, $\alpha(\vecx)$ is used to track the phases, with $\alpha=0.5$ indicating the interphase between the phases.  In this way, we can model the multiphase medium as a single fluid with variable density and viscosity as follows:
\begin{subequations}
\label{eq3}
\begin{align}
\rho(\vecx) = \rho_L\alpha(\vecx) + \left[1-\alpha(\vecx)\right]\rho_G, \\
\mu(\vecx) = \mu_L\alpha(\vecx) + \left[1-\alpha(\vecx)\right]\mu_G.
\end{align}
\end{subequations}
Here, $\rho_L$ and $\mu_L$ are the constant liquid density and viscosity, and similarly for $\rho_G$ and $\mu_G$ in the gas.

In this way, we can write the Navier--Stokes equations in a single-fluid formulation as follows:
\begin{subequations}
    \begin{equation}
\label{eq:ns-mom}
\rho(\vecx) \left( \frac{\partial \vecu}{\partial t} + \vecu \cdot \nabla \vecu \right) 
= -\vec{\nabla} p + \vec{\nabla} \cdot \left[ \mu(\vecx) \left( \nabla  \vecu + \nabla  \vecu^T \right) \right] 
- \rho(\vecx) g \hatz + \vecF_{ST},
\end{equation}
\begin{equation}
\label{eq:ns-incomp}
\nabla  \cdot \vecu = 0.
\end{equation}%
\label{eq:ns}%
\end{subequations}%
Here, $-\rho(\vecx)g\hatz$ is the body force due to gravity, pointing in the negative $z$-direction.  Furthermore, $\vecF_{ST}$ is the approximation to the surface-tension force in the VoF formulation, and is given by~\cite{brackbill1992continuum}:
\begin{equation}
\label{eq7}
\vecF_{ST} = -\surften\nabla \alpha \left[ \nabla \cdot \left(\frac{\nabla  \alpha}{\lvert \nabla  \alpha \rvert} \right) \right].
\end{equation}
The rationale behind this approximation is as follows.  In a two-fluid formulation of the Navier--Stokes equations, the surface tension is represented either by interfacial matching conditions across the phases or equivalently, by a force localized to the interface and hence, expressed in terms of a delta function.  In the latter formulation, one would have:
\begin{equation}
\vecF_{ST} = \surften \kappa \hatn \delta(\vecx-\vecx_I),
\label{eq:st1}
\end{equation}
where $\vecx_I$ denotes the interface location.  Since $\alpha=1/2$ in the volume-of-fluid formulation describes the interface location, we have:
\begin{equation}
\delta(\vecx-\vecx_I)=|\nabla\alpha|\delta\left(\alpha-\tfrac{1}{2}\right).
\label{eq:st2}
\end{equation}
Hence:
\begin{equation}
\hatn\,\delta(\vecx-\vecx_I)=\nabla\alpha\,\delta\left(\alpha-\tfrac{1}{2}\right).
\label{eq:st3}
\end{equation}
Furthermore, $\kappa=-\left[\nabla\cdot \left(\nabla\alpha/|\nabla\alpha|\right)\right]_{\alpha=1/2}$, hence:
\begin{equation}
\vecF_{ST}=-\surften \nabla\cdot\left(\frac{\nabla\alpha}{|\nabla\alpha|}\right) \nabla\alpha \,\delta\left(\alpha-\tfrac{1}{2}\right).
\label{eq:st4}
\end{equation}
A final step in the VoF approximation of the surface-tension is to `smear' the force over the entire fluid volume by omitting the delta function in Equation~\eqref{eq:st3}.  In this way, Equation~\eqref{eq7} is recovered.

The smearing of the surface-tension force over the entire fluid volume is justified because $\alpha$ is approximately constant away from the interface, meaning that the expression $\vecF_{ST}$ in Equation~\eqref{eq7} becomes negligible far from the interface.  However, this approximation can in some applications introduce `spurious currents' into the numerical model~\cite{solomenko2017mass}, which can be mitigated by a careful discretization of the Navier--Stokes equations, and by carrying out simulations at high resolution.

Finally, conservation of mass of the separate phases requires that:
\begin{equation}
\label{eq8}
\frac{\partial \alpha}{\partial t}+\nabla\cdot\left(\vecu \alpha\right)=0.
\end{equation}
Since the flow is incompressible, this is equivalent to:
\begin{equation}
\label{eq9}
\frac{\partial \alpha}{\partial t}+\vecu\cdot\nabla\alpha=0.
\end{equation}
To summarize, the relevant  equations of motion to be solved are:  Equations~\eqref{eq:ns-mom}--\eqref{eq:ns-incomp}, and Equation~\eqref{eq9}.  These are highly complicated coupled partial differential equations.  To solve these efficiently, we use the OpenFOAM software toolbox.

\subsection{OpenFOAM}

OpenFOAM  is a C++ toolbox for the development of customized numerical solvers for continuum mechanics problems, in particular Computational Fluid Dynamics (CFD). 
However, the toolbox comes equipped with a range of already-built solvers and tutorials, meaning that simple test-case simulations can be set up with reduced effort (i.e. compared to developing custom-built solvers).  One such solver is \texttt{interFOAM}, which enables one to simulate two-phase flow problems based on Equations~\eqref{eq:ns} and~\eqref{eq9}.

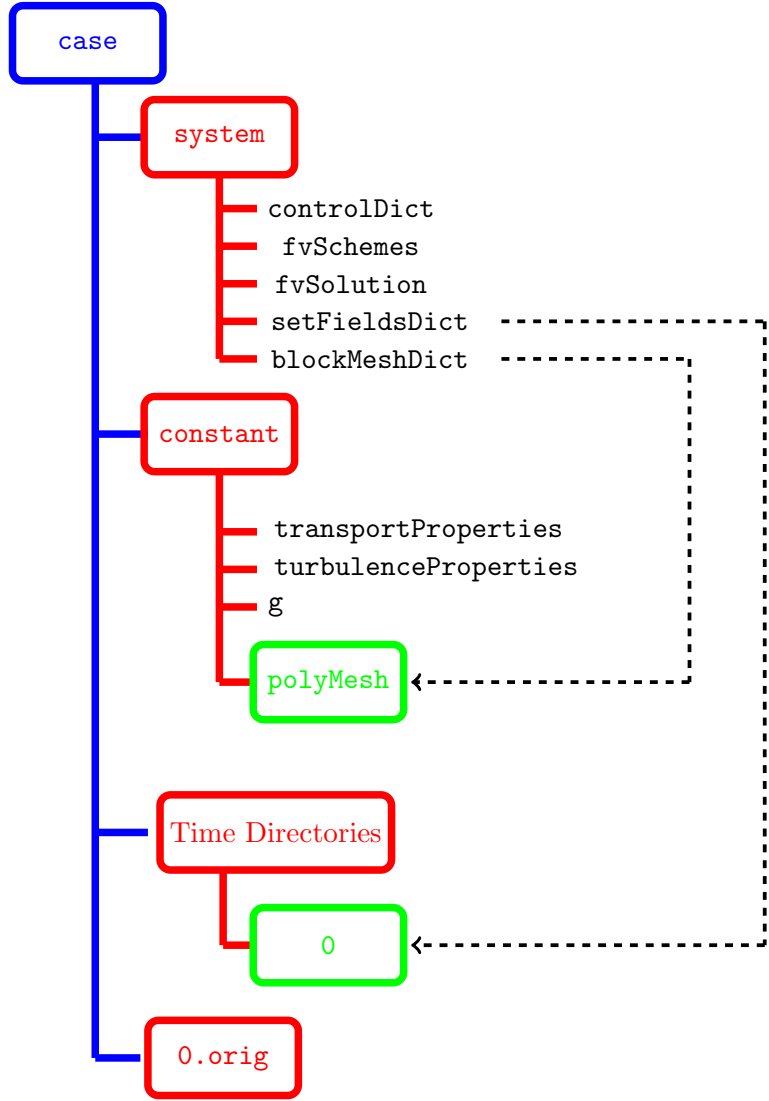
\begin{figure}
\centering
\begin{tikzpicture}
    \node[draw,color=blue,line width=1mm,rounded corners,minimum width=2cm, minimum height=1cm] at (0, 0.5)       {\texttt{case}};
    \draw [line width=1mm,color=blue] (0.1,0) -- (0.1,-13);
		\draw [line width=1mm,color=blue] (0.1,-0.75) -- (0.75,-0.75);
		\node[draw,color=red, line width=1mm,rounded corners,minimum width=2cm, minimum height=1cm] at (1.75, -0.75)       {\texttt{system}};
		\draw[line width=1mm,color=red] (1.75,-1.05-0.2) -- (1.75,-3.5-0.2);
		\draw[line width=1mm,color=red] (1.75,-1.5-0.2) -- (2.25,-1.5-0.2);
		\node at (3.5,-1.5-0.2) {\texttt{controlDict}};
		\draw[line width=1mm,color=red] (1.75,-2-0.2) -- (2.25,-2-0.2);
		\node at (3.5,-2-0.2) {\texttt{fvSchemes}};
		\draw[line width=1mm,color=red] (1.75,-2.5-0.2) -- (2.25,-2.5-0.2);
		\node at (3.5,-2.5-0.2) {\texttt{fvSolution}};
		\draw[line width=1mm,color=red] (1.75,-3-0.2) -- (2.25,-3-0.2);
		\node at (3.75,-3-0.2) {\texttt{setFieldsDict}};
		\draw[line width=0.5 mm, dashed] (5.5,-3-0.2) -- (9,-3-0.2);
		\draw[line width=0.5 mm, dashed] (9,-3-0.2) -- (9,-10-1.5);
		\draw[->,line width=0.5 mm, dashed] (9,-10-1.5) -- (4.3,-10-1.5);
		\draw[line width=1mm,color=red] (1.75,-3.5-0.2) -- (2.25,-3.5-0.2);
		\node at (3.75,-3.5-0.2) {\texttt{blockMeshDict}};
		\draw[line width=0.5 mm, dashed] (5.5,-3.5-0.2) -- (8,-3.5-0.2);
		\draw[line width=0.5 mm, dashed] (8,-3.5-0.2) -- (8,-7.5-0.5);
		\draw[->,line width=0.5 mm, dashed] (8,-7.5-0.5) -- (4.3,-7.5-0.5);
		\draw[line width=1mm,color=blue] (0.1,-4.5-0.2) -- (0.8,-4.5-0.2);
		\node[draw,color=red,line width=1mm,rounded corners,minimum width=2cm, minimum height=1cm] at (1.75, -4.5-0.2)       {\texttt{constant}};
		\draw[line width=1mm,color=red] (1.75,-4.73-0.5) -- (1.75,-7.5-0.5);
		\draw[line width=1mm,color=red] (1.75,-5.5-0.5) -- (2.25,-5.5-0.5);
		\node at (4.4,-5.5-0.5) {\texttt{transportProperties}};
		\draw[line width=1mm,color=red] (1.75,-6-0.5) -- (2.25,-6-0.5);
		\node at (4.5,-6-0.5) {\texttt{turbulenceProperties}};
		\draw[line width=1mm,color=red] (1.75,-6.5-0.5) -- (2.25,-6.5-0.5);
		\node at (2.5,-6.5-0.5) {\texttt{g}};
		\draw[line width=1mm,color=red] (1.75,-7.5-0.5) -- (2.25,-7.5-0.5);
		\node[draw,color=green, line width=1mm,rounded corners,minimum width=2cm, minimum height=1cm] at (3.2, -7.5-0.5)       {\texttt{polyMesh}};
		\draw[line width=1mm,color=blue] (0.1,-9-1) -- (0.8,-9-1);
		\node[draw,color=red,line width=1mm,rounded corners,minimum width=2cm, minimum height=1cm] at (2.5, -9-1)       {Time Directories};
		\draw[line width=1mm,color=red] (1.8,-9.3-1.2) -- (1.8,-10-1.5);
		\draw[line width=1mm,color=red] (1.8,-10-1.5) -- (2.2,-10-1.5);
		\node[draw,color=green,line width=1mm,rounded corners,minimum width=2cm, minimum height=1cm] at (3.2, -10-1.5)       {\texttt{0}};
		\draw[line width=1mm,color=blue] (0.1,-13) -- (0.7,-13);
		\node[draw,color=red,line width=1mm,rounded corners,minimum width=2cm, minimum height=1cm] at (1.8, -13)       {\texttt{0.orig}};
\end{tikzpicture}
\caption{Directory structure for a typical OpenFOAM simulation.  The broken lines with arrowheads show the effect of running the \texttt{setFields} and \texttt{blockMesh} commands on various directories.}
\label{fig:directory}
\end{figure}

The key elements of an OpenFOAM transient CFD simulation can be explained by reference to the directory structure of a sample simulation with directory name \texttt{case}, shown in Figure~\ref{fig:directory}.
We show this here because it paves the way for subsequent discussion on the treatment of boundary and initial conditions.  For the same purpose, we summarize the contents of the sub-directories as follows:
\begin{enumerate}[noitemsep]
\item Sub-directory \texttt{system} contains the following files:
	\begin{enumerate}[noitemsep]
		\item \texttt{controlDict}: Simulation parameters are set, e.g. the end-time of the simulation, the frequency at which simulation data is outputted to files.
		\item \texttt{fvSchemes}: The algorithms for the solution Equations~\eqref{eq:ns} and~\eqref{eq9} are chosen.
		\item \texttt{fvSolution}: Further algorithms for the solution Equations~\eqref{eq:ns} and~\eqref{eq9} are chosen (e.g. the time-marching scheme).
		\item \texttt{setFieldsDict}: Initial conditions on $\alpha$, $\vecu$, and $p$ are prescribed.
		\item \texttt{blockMeshDict}: Specifications for the computational domain are given.  The geometry of the fluid
		domain is described, together with the grid resolution and any local grid refinement that is required.   Boundary conditions on the various faces making up the domain are also supplied.    Once the mesh parameters are specified, a mesh is generated using the \texttt{blockMesh} command.  The resulting mesh is then copied into the directory \texttt{polyMesh}.  All domain dimensions are given in metres.
		\end{enumerate}
\item Sub-directory \texttt{constant} contains the following files:
	\begin{enumerate}[noitemsep]
		\item \texttt{transportProperties}:  The surface tension, liquid and gas densities and viscosity ate set using S.I. units.
		\item \texttt{turbulenceProperties}: Turbulence modelling, if required, is specified.
		\item \texttt{g}: The magnitude and direction of the gravity vector is specified.
		\end{enumerate}
\item Time Directories:
\begin{enumerate}[noitemsep]
    \item \texttt{0}: An initial configuration is copied from \texttt{0.orig} into \texttt{0}.  The initial configuration is over-written with custom initial conditions using the \texttt{setFields} command.  This sets the state of the system ($\alpha$, $\vecu$, and $p$) at $t=0$.  The original initial conditions are kept always in \texttt{0.orig} so that the simulation can be recreated from scratch, if needs be.
		\item Further directories corresponding to the state of the system ($\alpha$, $\vecu$ and $p$) at later times $t_1,t_2,\dots$ are created here when the simulation is run.  The values $t_1,t_2,\cdots$ are set in \texttt{controlDict}.
		\end{enumerate}
\end{enumerate}
A simulation such as the one outlined here is executed in OpenFOAM using the \texttt{interFOAM} command.  We do not elaborate further on the use of \texttt{interFOAM} here as our main purpose here is to define reference points for subsequent discussion of initial and boundary conditions.  Instead, the interested reader is referred to the rich online resources on OpenFOAM for further instruction (e.g. Reference~\cite{nagyInter}, specifically on \texttt{interFOAM}).

\subsection{olaFLOW}

\begin{figure}
\centering
\begin{tikzpicture}
\draw [line width=0.5mm] (0,0) -- (10,2);
\draw [line width=0.5mm] (0,0) -- (0,1);
\draw [line width=0.5mm] (0,1) -- (10,3);
\draw [line width=0.5mm] (10,2) -- (10,3);
%
%
\draw [line width=0.25mm,color=blue] (0,0+0.5) -- (10,2+0.5);
\draw [line width=0.25mm,color=blue] (10,2.5) -- (10-0.5,2.5+0.25);
\draw [line width=0.25mm,color=blue] (10-0.5,2.5+0.25)--(-0.5,0.75);
\draw [line width=0.25mm,color=blue] (-0.5,0.75)--(0,0.5);
\draw [<->,line width=0.25mm] (-0.5-0.25,0.25) -- (-0.5-0.25,0.25+0.5);
\node at (-0.5-1.5,0.5) [draw,fill=white] {$h=0.05\,\mathrm{m}$};
\draw [line width=0.5mm] (0,0) -- (-0.5,0.25);
\draw [line width=0.5mm] (-0.5,0.25) -- (-0.5,1.25);
\draw [line width=0.5mm] (-0.5,1.25) -- (0,1);
\draw [line width=0.5mm] (-0.5,1.25) -- (-0.5+10,1.25+2);
\draw [line width=0.5mm] (-0.5+10,1.25+2) -- (10,3) ;
\draw [line width=0.5mm,dashed] (10,2) -- (-0.5+10,1.25+2-1);
\draw [line width=0.5mm,dashed] (-0.5+10,1.25+2) -- (-0.5+10,1.25+2-1);
\draw [line width=0.5mm,dashed] (-0.5+10,1.25+2-1) -- (-0.5,0.25);
\draw [<->,line width=0.25mm] (0,0-0.5) -- (10,2-0.5);
\node at (5,0.45) [draw,fill=white] {$L_x=1\,\mathrm{m}$};
\draw [-,line width=0.25mm,dashed] (0,0) -- (0,-0.5);
\draw [-,line width=0.25mm,dashed] (-0.5,0.25) -- (-0.5,-0.25-0.1);
\draw [<->,line width=0.25mm] (0,0-0.5) -- (-0.5,0.25-0.5);
\node at (-0.6,-0.8) [draw,fill=white] {$L_y$};
\draw [<->,line width=0.25mm] (10.25,2) -- (10.25,3);
\node at (11.45,2.5) [draw,fill=white] {$L_z=0.1\,\mathrm{m}$};
\draw [->,line width=0.5mm,color=blue] (10-1,0-1) -- (11-1,0.25-1);
\node at (11-1,0.25-1+0.25)  {{\textcolor{blue}{$x$}}};
\draw [->,line width=0.5mm,color=blue] (10-1,0-1) -- (10-1,1.1-1);
\node at (10-1,1.1-1+0.25)  {{\textcolor{blue}{$z$}}};
\draw [->,line width=0.5mm,color=blue] (10-1,0-1) -- (10-0.85-1,0.37-1);
\node at (10-0.85-1,0.37-1+0.25)  {{\textcolor{blue}{$y$}}};
%
%
\fill[blue!20,opacity=0.7] 
    (0,0) -- (10,2) -- (10,2.5) -- (10-0.5,2.75) -- (-0.5,0.75) -- (0,0.5) -- cycle;
\fill[blue!20,opacity=0.7] 
		(0,0) -- (0,0.5) -- (-0.5,0.75) -- (-0.5 , 0.25) -- cycle;
\end{tikzpicture}
\caption{Schematic diagram showing the geometry of the simulation, together with the initial condition for the water level.  The unit normal $\hat{\mathbf{y}}$ referred to in the text points in the positive $y$-direction.}
\label{fig:blockMesh_schematic}
\end{figure}
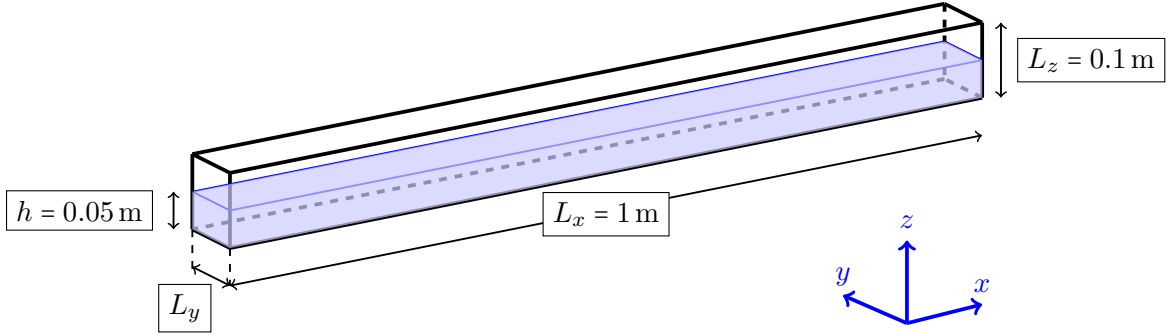

We set up a simulation in a simple rectangular domain such as the one shown in Figure~\ref{fig:blockMesh_schematic}.  We apply standard no-slip conditions at the bottom wall and standard atmospheric conditions at the top of the domain.  Simple `empty' boundary conditions are applied in the faces whose normals are $\pm \hat{\mathbf{y}}$, these enforce a two-dimensional flow.  
A simple uniform mesh is created based on this domain using the \texttt{blockMesh} command in OpenFOAM.  However, we encounter difficulty when we seek to apply a time-varying boundary condition at the inlet and a wave-damping boundary condition at the outlet.

Rather than writing OpenFOAM code from scratch to describe an oscillating inlet, we resort to a suite of OpenFOAM cases called olaFlow which has been developed to simulate flows in wave tanks using the Volume-of-Fluid methodology.  The olaFlow codes are available on an online repository~\cite{olaFlow} and a simple tutorial corresponding to a two-dimensional piston wavemaker in a flume is already set up (specifically, \texttt{wavemakerFlume}).  
Detailed instructions on how to download  and execute olaFlow are provided~\ref{sec:app:ola}.  
To model the tabletop flume, we use  the tutorial case \texttt{wavemakerFlume} with our own custom initial conditions and geometry, which we describe as follows.
\begin{itemize}[noitemsep]
\item {\textbf{Geometry:}}  The geometry of the simulation shown in Figure~\ref{fig:blockMesh_schematic}, to match the tabletop flume shown in Figures~\ref{fig:wt1}--\ref{fig:wt3}.  We use $L_x=1.0\,\mathrm{m}$ to match the tabletop flume, as well as $h=0.05\,\mathrm{m}$.  The domain in the $z$-direction is chosen to be $L_z=10\,\mathrm{cm}$, comprising liquid in the bottom half and gas (air) in the top half.  We also use $L_y=2\,\mathrm{cm}$ in the $y$-direction.  Although this $L_y$-value does not match the tabletop flume, this is not important, as we seek to simulate a fundamentally two-dimensional wave problem.  This choice of narrow channel width in the simulation, together with `empty' boundary conditions enforces a two-dimensional flow.  
\item {\textbf{Mesh Resolution:}}  The computational mesh is a uniform one, with $500$ cells in the $x$-direction, $50$ cells in the $z$-direction, and $1$ cell in the $y$-direction, corresponding to a two-dimensional flume.  We have verified that this resolution is adequate by performing a mesh-refinement study (below).
\item {\textbf{Initial Conditions:}} These are zero velocity in the water and in the air, and a flat interface, corresponding to $\alpha=1$ in the liquid and $\alpha=0$ in the water.  
The initial conditions are specified in the \texttt{0.orig} directory in Figure~\ref{fig:directory} and in the \texttt{setFields} dictionary.   
\item{\textbf{Boundary Conditions:}} Boundary conditions are specified in the $\texttt{0.orig}$ directory and in the \texttt{blockMeshDict} dictionary.  By using olaFlow, we are able to set up the inlet as a boundary wall which oscillates back and forth with a set frequency.  The oscillating inlet is implemented in the olaFlow code \texttt{pistonWaveGen.py}.  We have modified this code so that the inlet location $x=\xi$ is a simple function of time,
\begin{equation}
\xi(t)=(S/2)\cos(\omega t),
\end{equation}
where $S$ and $\omega$ are set by the user.  This enables us to create a simulation which mimics the tabletop flume as closely as possible.  Finally, the outlet wall is set up so as to absorb outgoing waves, using the olaFlow outlet boundary type \texttt{waveAbsorption2DVelocity} as boundary condition for the velocity.  This absorbing boundary condition is described in the PhD thesis of Higuera~\cite{caubilla2015aplicacion} which provides the theoretical underpinning for olaFlow.
\end{itemize}

\subsection{Mesh-Refinement Study}

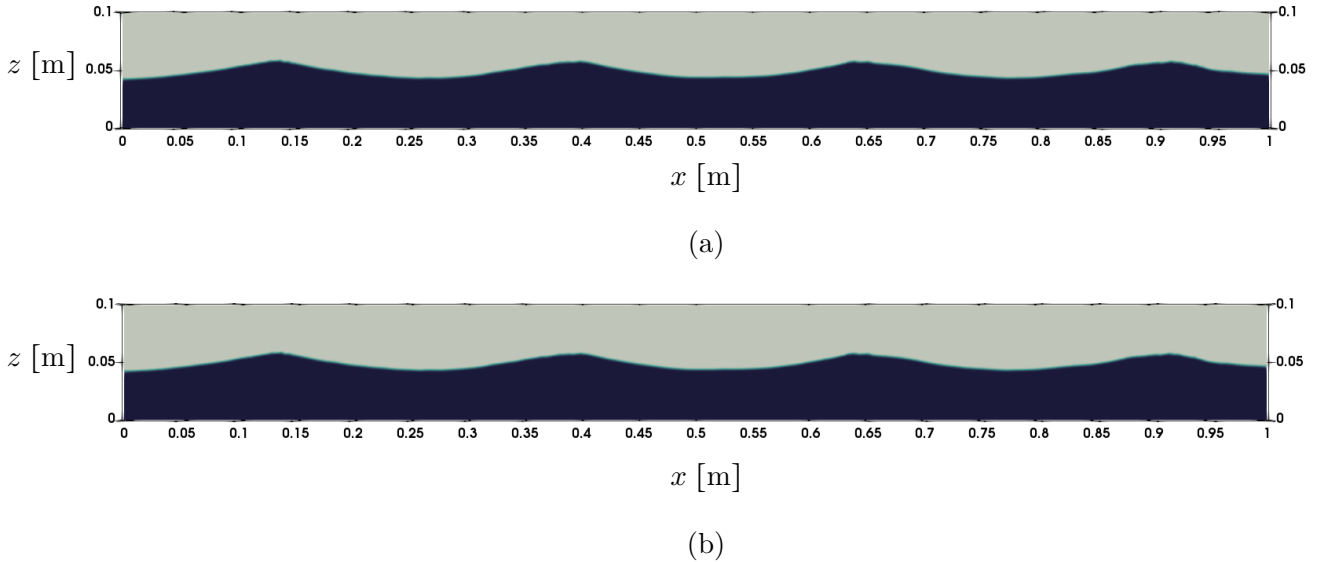
\begin{figure}[htb]
\centering
\begin{tikzpicture}
		\node[anchor=south west] (img1) at (0,0) {\includegraphics[width=17cm]{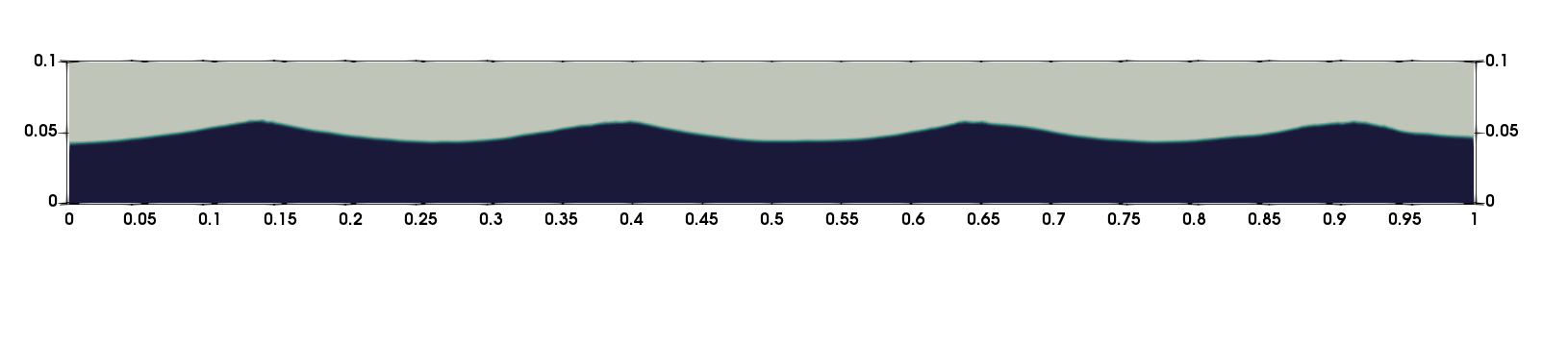}};
    \node[anchor=south] at ([xshift=0mm,yshift=-2mm]img1.south) {(a)};
		\node[anchor=west] at ([xshift=-8mm,yshift=5mm]img1.west) {$z\,\left[\mathrm{m}\right]$};
		\node[anchor=south] at ([xshift=0mm,yshift=7mm]img1.south) {$x\,\left[\mathrm{m}\right]$};

		\node[anchor=south west] (img2) at (0,-4) {\includegraphics[width=17cm]{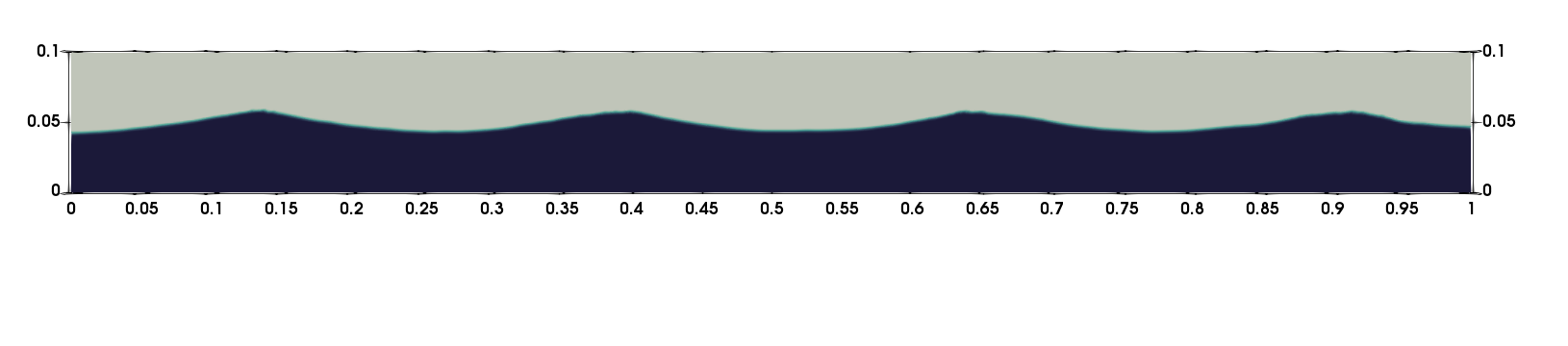}};
    \node[anchor=south] at ([xshift=0,yshift=-2mm]img2.south) {(b)};
		\node[anchor=west] at ([xshift=-8mm,yshift=6mm]img2.west)  {$z\,\left[\mathrm{m}\right]$};
		\node[anchor=south] at ([xshift=0mm,yshift=7mm]img2.south) {$x\,\left[\mathrm{m}\right]$};
		
\end{tikzpicture}
\caption{Mesh-refinement study: snapshot ($xz$-plane) of the free surface at $t=8.5\,\mathrm{s}$.  Top: standard mesh $(N_x,N_y,N_z)=(500,1,50)$.   Bottom: fine mesh $(N_x,N_y,N_z)=(1000,1,100)$.  Simulation parameters: $S=1.1\,\mathrm{cm}$, $\omega=138\,\mathrm{RPM}$.}
\label{fig:refinement}
\end{figure}

For the present purposes, the reference mesh is uniform blockmesh of size $(N_x,N_y,N_z)=(500,1,50)$.  Here, we report very briefly on simulation results where the blockmesh has been increased to $(1000,1,100)$.   We carry out simulations for a stroke length $S=0.011\,\mathrm{m}$ and a forcing period $T=0.435\,\mathrm{s}$ (hence, $\omega=138\,\mathrm{RPM}$).
A comparison between the standard mesh and the fine mesh is shown in Figure~\ref{fig:refinement}.  
There is no difference between the two results, confirming that the simulation can be deemed to have converged when using the reference mesh.

\subsection{First results}

\begin{figure}[htb]
	\centering
		\includegraphics[width=0.98\textwidth]{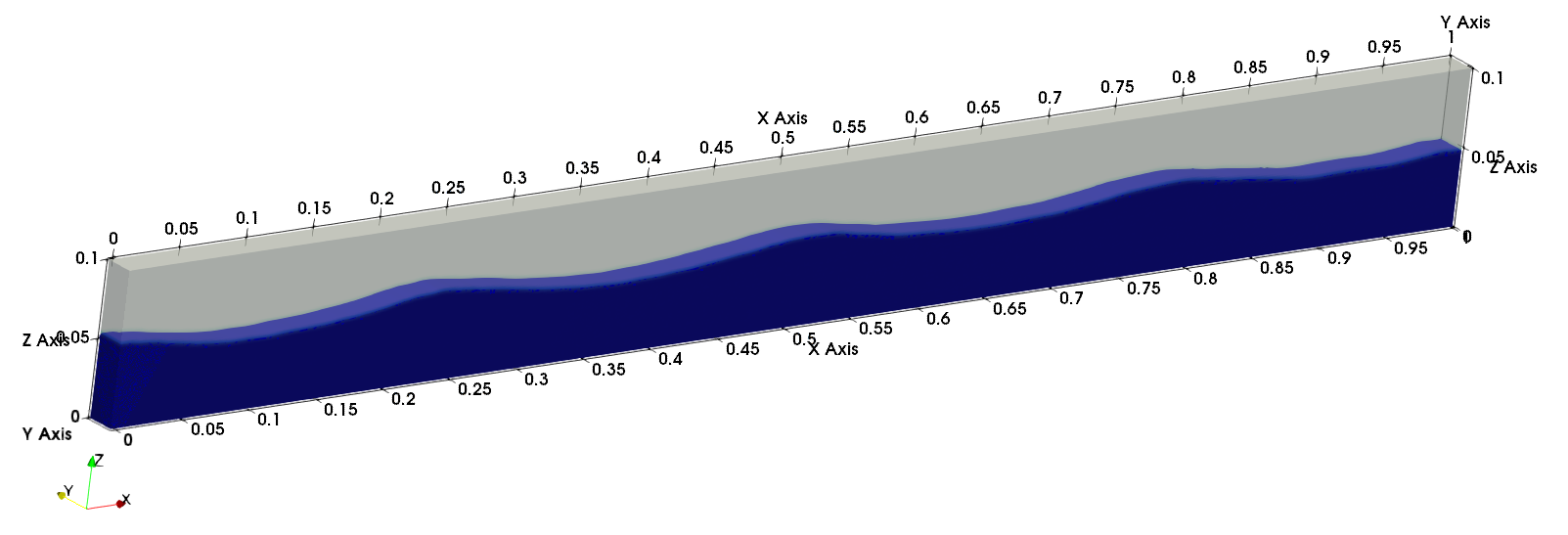}
		\caption{Representative result: snapshot of the free surface at $t=8.5\,\mathrm{s}$.    Simulation parameters: $S=1.0\,\mathrm{cm}$, $\omega=133\,\mathrm{RPM}$.  Axes scales in metres.}
	\label{fig:screenshot_t8p5}
\end{figure}

Representative results results are shown in Figure~\ref{fig:screenshot_t8p5}, for a test case with $T=0.45\,\mathrm{s}$ (hence, $\omega=133\,\mathrm{RPM}$, close  to the value in the experiment in Section~\ref{sec:flume}), and $S=0.01\,\mathrm{m}$.  The results represent a snapshot taken at
 $t=8.5\,\mathrm{s}$, such that the piston wavemaker has had sufficient time to produce a fully-developed wave train.  Results have been visualized with Paraview, which interfaces very easily with OpenFOAM.

\begin{figure}[htb]
	\centering
		\includegraphics[width=0.8\textwidth]{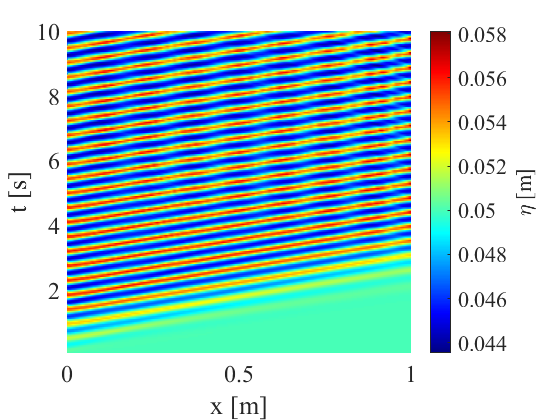}
		\caption{Spacetime plot of the free surface $\eta(x,t)$ (simulation results).  Simulation parameters: $S=1.0\,\mathrm{cm}$, $\omega=133\,\mathrm{RPM}$.}
	\label{fig:spacetime_numerics}
\end{figure}

A single monochromatic wave can be identified in  Figure~\ref{fig:screenshot_t8p5}. 
We can visualize this more precisely  by taking a two-dimensional planar slice.  The plane has a normal vector in the $y$-direction and is centred at $L_y/2$.  By extracting the free surface at each point in time, and plotting the result in a space-time plot, we confirm that the monochromatic wave is a travelling wave (Figure~\ref{fig:spacetime_numerics}).

\begin{remark}
In this section we shift the origin in the $z$-direction, such that the mean water level is at $z=h_0$.  Thus, the free surface is represented by $\eta(x,t)=h_0+[\text{Disturbance}]$.
\end{remark}

\begin{figure}[htb]
	\centering
		\includegraphics[width=0.7\textwidth]{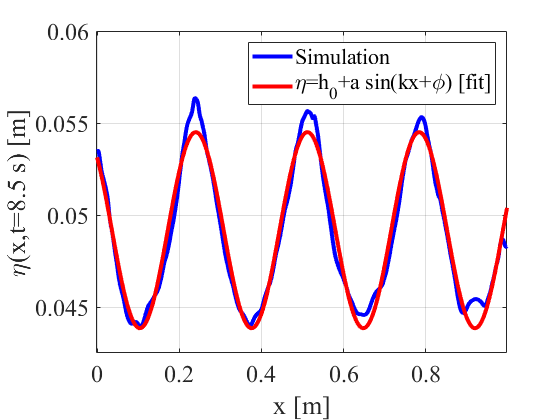}
		\caption{Snapshot of the free-surface height at $t=8.5\,\mathrm{s}$.  Simulation parameters: $S=1.0\,\mathrm{cm}$, $\omega=133\,\mathrm{RPM}$.}
	\label{fig:get_k_dns}
\end{figure}

To analyze the travelling wave more precisely, we again focus the snapshot at $t=8.5\,\mathrm{s}$, well after transient effects have died away.  A planar slice of the snapshot is shown in Figure~\ref{fig:get_k_dns}.
  We have fitted a sinusoidal curve to the snapshot, $\eta(x)=h_0+A\sin(kx+\varphi)$, and the result is $k= 23.12\,\mathrm{m}^{-1}$ and $A=0.0066\,\mathrm{m}$, corresponding to $\lambda=0.272\,\mathrm{m}$.
  The predicted value of the wavelength according to linear theory (\textit{cf}. Figure~\ref{fig:DR_sample}, $\omega=2\pi/0.45\,\mathrm{rad}\cdot\mathrm{s}^{-1}$) is $\lambda=0.265\,\mathrm{m}$,  Furthermore, the measured value of  $H/S$ is $H/S=2a/S=1.07$.  This is to be compared with the linear theory (\textit{cf}. Equation~\eqref{eq:HoS}), which predicts $H/S=1.15$ at $\lambda=0.265\,\mathrm{m}$.
The agreement here between the numerical simulations and the linear theory can be considered very close.  Discrepancies between the simulations and the theory can be accounted for by the fact that the simulated wave is not perfectly monochromatic: secondary oscillations at shorter wavelength can be seen in Figure~\ref{fig:get_k_dns}, especially at wave crests and troughs.  This can be explained by nonlinear effects.  We address this in more detail below.

%
%

\subsection{Height-to-stroke ratio}

\begin{figure}[htb]
\centering
\begin{tikzpicture}[scale=1.2,transform shape]
    \begin{axis}[
        xlabel={$S\,\,\mathrm{mm}$},
        ylabel={$H\,\,\mathrm{mm}$},
        grid=both,
        width=10cm,
        height=8cm,
        thick,
        mark options={solid},
				ymin=6, 
        ymax=20, 
				xmin=5, 
        xmax=20, 
        legend style={at={(0.5,-0.2)},anchor=north,legend columns=1}
    ]
		
		    \addplot[
				only marks, 
        mark=o,
        color=red
    ] coordinates {
		    (6, 6.4)
        (7, 7.6)
				(8, 8.8)
				(9, 9.669)
				(10,10.65)
				(11, 11.592)
				(12, 12.6307)
				(13, 13.4436)
				(14, 14.342)
				(15, 14.594797)
    };
		
    \addplot[
        domain=6:20, 
        samples=100,
        dashed,
        thick,
        color=red
    ] {1.1501*x};
		
    \addplot[
        only marks,
        mark=*,
        thick,
        color=red
    ] coordinates {
        (15, 14.594797)
    };
		\node[anchor=south west, inner sep=0]  at (75,2) {
        \includegraphics[width=0.25\textwidth]{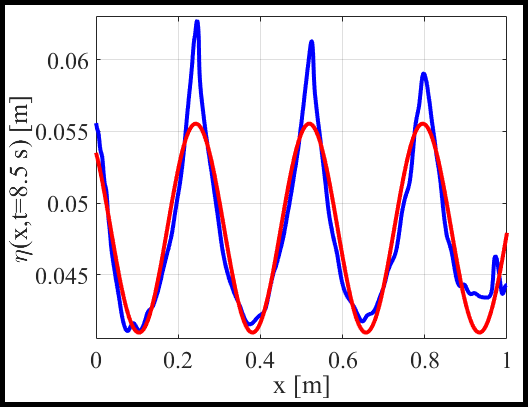}
    };
		\draw[->,red] (100,82)--(100,72);
    \end{axis}
\end{tikzpicture}
\caption{Dependence of wave height $H$ on the stroke length $S$ of the piston wavemaker.  Circles: numerical simulations.   Red line: $H/S=1.15$, from linear theory, with $\lambda=0.265\,\mathrm{m}$.  The filled data-point corresponds to $S=15\,\mathrm{mm}$, when the linear theory clearly breaks down.  The inset shows a snapshot of the free-surface height at the same stroke length, and exhibits significant wave steepening. }
\label{fig:stroke_study}
\end{figure}

We have extended the study in the previous section to a range of different stroke lengths $S$.  We have thereby computed the dependence of the wave height on $S$.
The results are summarized in Figure~\ref{fig:stroke_study}.  For small values of $S$, there is good agreement between the linear theory (broken red line) and the simulations.  At $S\approx 14\,\mathrm{mm}$, significant wave steepening occurs, and the linear theory breaks down.    Beyond this point, the waves steepen significantly, as seen in the inset in Figure~\ref{fig:stroke_study}.

%
%

\section{Student Experiences}
\label{sec:feedback}

The theoretical modelling of linear water waves (similar to Section~\ref{sec:theory}--\ref{sec:theory_closed}) was introduced to students in ACM 40890 \textit{Advanced Fluid Mechanics} at University College during the 2024–2025 academic year.
This is an advanced module, taken by final-year Bachelor of Science students majoring in Theoretical Physics and Applied Mathematics, as well as by first-year PhD students.  Ten students took the module in 2024-2025, in the Spring Trimester (12 weeks). 
Alongside lectures on theory, students completed a substantial assignment involving an experiment with the tabletop flume, as well as CFD simulations of same, using olaFlow.  

To assess student perspectives on this integrated approach, a survey was administered in April 2025 via an anonymous Google Form, following approval from the University's Human Research Ethics Committee.  Participation was voluntary; 5 anonymous responses were obtained.   While the sample size is too small for statistical analysis, the feedback offers useful qualitative insights for instructors considering similar module designs.

Overall, students reported that the integration of theory, experiment, and simulation enhanced their understanding of complex fluid mechanics concepts. One student remarked:
\begin{quote}
My perception of experiments in Fluid Mechanics is they are insightful and provide an opportunity to solidify and gain a greater understanding the various topics we studied in Fluid Mechanics. It gives an opportunity to tweak variables and see how they affect the outcome and gain a more intuitive understanding of many phenomena than studying the Mathematics alone does not achieve.
\end{quote}
Students also highlighted the value of simulations in bridging theory and experiment:
\begin{quote}
I learned a lot from how we used the data from the experiments to help us set up the simulations and compare the theory, experiments, and simulations. As someone who never really liked experiments and always liked simulations I never really saw the connection between the two until now. I found great learning in that.
\end{quote}

On a cautionary note, the students noted the steep learning curve involved in setting up the CFD simulations, particularly given the lack of dedicated technical support. One student reflected:
\begin{quote}
CFD simulations require a broad skill-set; learning to use various applications (many of which lack documentation), programming, data analysis of large data sets, and understanding of the underlying workings of algorithms to understand when/why it does not behave the way you expect and if it is a syntax issue, physics issue, a known limitation or a bug. It is a steep learning curve and working through these issues comes comes with learning and you usually don't make the same error twice. 
Though  to make my point, I found various other hurdles in conducting CFD simulations limited the ability to focus to the Fluids Mechanics of the scenarios being studied.
\end{quote}
This was echoed by other students in offline feedback.  Overall, the students felt that more formal lectures in CFD (involving detailed cases studies) would be beneficial, as opposed to relying on a mixture of theory-heavy face-to-face lectures and posted online CFD tutorials.

Finally, the instructor observed in the completed reports on the project that while students proved adept at extracting summary information from the experimental data, few were able to see beyond the summary information, and to go down to the level of detail required for a complete understanding of the observed wave forms.  While some of this must certainly be due to time constraints on the students' part, in future, the instructor should place more emphasis on data analysis techniques, to help students develop deeper insights from experimental results.

\section{Conclusions}
\label{sec:conc}

Summarizing, we have described in detail the theory of small-amplitude water waves in the limit where the equations of motion can be linearized.  The theory is presented is spatio-temporal, as it describes the response of the system in space and time to localized inlet forcing.  As a by-product the classical dispersion relation for linear water waves (Equation~\eqref{eq:disp1}) is recovered.

As part of an integrated approach to instruction in the topic, we have developed -- and present here -- an inexpensive $1\,\mathrm{m}$-long tabletop flume and variable-RPM Lego wavemaker, which can be used to gather experimental data.  With the advent of high-quality mobile-phone cameras such data can be readily recorded and analysed.  Hence, we report  on a representative experiment.  Using  nonlinear least squares fitting to perform a statistical analysis on the data, we have found excellent agreement between the theory and the experiments, once the presence of both travelling waves and standing waves is accounted for.  Future student work on the flume and the wave-maker could focus on the boundary conditions -- for instance, on engineering a purely reflective boundary condition, which would also open up the possibility of finding resonant modes.

To complement the theoretical and experimental work, we have performed numerical simulations of the wave tank using OpenFOAM.
In the  simulations, it is much more straightforward to provide an absorbing boundary condition, and hence to produce a train of  clearly defined travelling waves, whose parameters agree with the predictions of the linear theory.  For small stroke lengths of the piston wavemaker, the results of the numerical simulations for  the height-to-stroke ratio are also in good agreement with the theory. 

Finally, we have reported also on the results of incorporating such an integrated approach -- involving theory, experiments, and numerical simulation in the classroom setting.  Feedback from the students showed that while students found the experience enhanced their learning, they also noted the need for better support in setting up and running CFD simulations.

\subsection*{Acknowledgments}

L\'ON  thanks Maria Meehan for advice on making a submission to the UCD Human Research Ethics Committee.  NY acknowledges support by the UCD School of Mathematics and Statistics through a funded summer research placement.

\appendix

\section{Installation of olaFlow}
\label{sec:app:ola}

In this work, we use the olaFlow to simulate the generation of water waves.  These are a suite of codes which can be run in the OpenFOAM computational framework.  Downloading and installing olaFlow is not trivial, as successful implementation of olaFlow relies on the use of an earlier version of OpenFOAM.  Hence, in this Appendix we outline the steps necessary to install olaFlow and execute simple test cases.  We describe this for users using a Windows Operating System, however, most of the commands for these tasks will be the same in Ubuntu.

First, we install an Ubuntu virtual machine which will run in a Windows environment, this is done using the Windows Power Shell and the command:
\begin{lstlisting}
wsl --install -d Ubuntu-18.04
\end{lstlisting}
We now initialize Ubuntu 18.04 by typing the following line into the Windows Power Shell terminal:
\begin{lstlisting}
wsl -d Ubuntu-18.04
\end{lstlisting}
Next, we install the appropriate version of OpenFOAM, which is OpenFOAM6.  The appropriate commands are obtained from \hyperlink{openfoam.org}{https://openfoam.org/download/6-ubuntu/} and are repeated here:
\begin{lstlisting}[breaklines=true,frame=single]
sudo sh -c "wget -O - https://dl.openfoam.org/gpg.key | apt-key add -"
sudo add-apt-repository http://dl.openfoam.org/ubuntu
\end{lstlisting}
Update the apt package list:
\begin{lstlisting}
sudo apt-get update
\end{lstlisting}
Install OpenFOAM6:
\begin{lstlisting}
sudo apt-get -y install openfoam6
\end{lstlisting}
Update the installation:
\begin{lstlisting}
sudo apt-get update
sudo apt-get upgrade
\end{lstlisting}
At the end of the installation, we must append  the following line to our \texttt{bashrc} file (located in \texttt{Ubuntu-18.04\textbackslash home\textbackslash username}):
\begin{lstlisting}
% . /opt/openfoam6/etc/bashrc
source /opt/openfoam6/etc/bashrc
\end{lstlisting}
We close the terminal and open a new one.  We then start a new virtual Ubuntu session.  By virtue of having modified the \texttt{bashrc}, OpenFoam 6 is automatically loaded.
%
%
%
%
This can be checked by typing:
\begin{lstlisting}
simpleFoam -help
\end{lstlisting}
We are now able to download olaFlow.  This can be done using the \texttt{clone} command in GitHub.  Alternatively, the installation can be downloaded as a zip file from:
\[
\texttt{\hyperlink{github}{https://github.com/phicau/olaFlow}}.
\]
This will produce the zip file \texttt{olaFlow-master.zip}, which we move in to the relevant directory and unzip.  We next modify the permissions on the unzipped folder:
\begin{lstlisting}
sudo chmod -R 777 olaFlow-master
\end{lstlisting}
We install `make':
\begin{lstlisting}
sudo apt-get install make
\end{lstlisting}
We can now use the OpenFOAM command `all make' to generate the appropriate olaFlow executables.  First:
\begin{lstlisting}
cd olaFlow-master
./allMake
\end{lstlisting}
Then:
\begin{lstlisting}
cd genAbs
./allMake
\end{lstlisting}
To run the various codes that generate transient boundary conditions in olaFlow, we require Python.  Hence, we install \texttt{pip}:
\begin{lstlisting}]
sudo apt install python-pip
\end{lstlisting}
and finally, \texttt{numpy}:
\begin{lstlisting}
sudo pip install numpy
\end{lstlisting}
Once these installations are complete, olaFlow is ready to be used.  For instance, to run a simple flume model from the tutorials, we change into the relevant directory:
\begin{lstlisting}
cd tutorials
cd wavemakerFlume
\end{lstlisting}
The relevant OpenFOAM case can be run using the appropriate file.  For instance, for the piston wave maker, we type:
\begin{lstlisting}
. runCasePiston
\end{lstlisting}
The details concerning this case (mesh generation, initial conditions, etc.) can be found by opening the file of the same name.
%


\end{document}